\documentclass[aip,graphicx]{revtex4-1}
\usepackage{graphicx}

\begin{document}


\title{Effective decrease of photoelectric emission threshold from gold plated surfaces} 



\author{Peter J Wass}
\altaffiliation{Department of Mechanical and Aerospace Engineering, MAE-A, P.O. Box 116250, University of Florida, Gainesville, Florida 32611, USA}
\author{Daniel Hollington}
\author{Timothy J Sumner\footnote{Corresponding author: t.sumner@imperial.ac.uk}}
\altaffiliation{Department of Physics, 2001 Museum Road, P.O. Box 118440, University of Florida, Gainesville, Florida 32611, USA}
\affiliation{High Energy Physics Group, Physics Department, Imperial College London, Blackett Laboratory, Prince Consort Road, London, SW7 2AZ, UK}
\author{Fangchao Yang}
\affiliation{Huazhong University of Science and Technology, Luoyu Road 1037, Wuhan, Hubei, 430074, China}
\author{Markus Pfeil}
\affiliation{University of Applied Sciences Ravensburg-Weingarten, Doggenriedstrasse, 88250 Weingarten, Germany}


\date{\today}

\begin{abstract}
Many applications require charge neutralisation of isolated test bodies and this has been
successfully done using photoelectric emission from surfaces which are electrically benign
(gold) or superconducting (niobium). Gold surfaces nominally have a high work function ($\sim 5.1$\,eV)
which should require deep UV photons for photoemission. In practice it
has been found that it can be achieved with somewhat lower energy photons with indicative work functions of ($ 4.1-4.3$\,eV). A detailed
working understanding of the process is lacking and this work reports on a study of the photoelectric
emission properties of $4.6\times4.6$\,cm$^2$ gold plated surfaces, representative of those used in typical satellite applications with a film thickness of 800\,nm, and measured surface roughnesses between 7 and 340\,nm. Various UV sources with photon energies from 4.8 to 6.2\,eV and power outputs from 1\,nW to 1000\,nW, illuminated a $\sim 0.3$\,cm$^2$ of the central surface region at angles of incidence from 0 to 60$^\circ$. 

Final extrinsic quantum yields in the range $10\,$ppm to $44\,$ppm were reliably obtained during 8 campaigns, covering a $\sim$3 year period, but with intermediate long-term variations lasting several weeks and, in some cases, bake-out procedures at up to 200$^{\mbox{o}}$\,C.  Experimental results were obtained in a vacuum system with a baseline pressure of $\sim 10^{-7}$\,mbar at room temperature.  A working model, designed to allow accurate simulation of any experimental configuration,  is proposed.
\end{abstract}


\maketitle 

\section{Introduction}
\label{intro}
Photoelectric emission is used for charge control of test bodies in a number of applications, both realised, such as Gravity Probe B (GP-B)~\cite{buchman95}, UV-LED~\cite{saraf16} and LISA Pathfinder (LISAPF)~\cite{armano16}, and proposed, such as STEP~\cite{sumner07}, LISA~\cite{sumner04, sun06, pollack10}, DECIGO~\cite{kawamura11}, BBO~\cite{crowder05}, TAIJI~\cite{hu17}, TianQin~\cite{luo16} and B-DECIGO~\cite{kawamura18}.  GP-B and LISAPF used $2537\, \rm{\AA}$ atomic line photons from a mercury vapour lamp, whilst UV-LED used light emitting diodes (LEDs) with a centre wavelength of 255nm and FWHM of 11nm.  GP-B used (superconducting) niobium plated surfaces (work function of 4.3\,eV) on the gyroscopes and gold plating, with a nominal work function of 5.1\,eV~\cite{kaye}, on the surrounding housing. UV-LED and LISAPF used gold plated surfaces on both test mass and housing. 254\,nm photons have an energy of 4.9\,eV which is usefully above the niobium work function but is nominally too low, to produce emission from gold. However GP-B, UV-LED and LISAPF were validated through empirical testing~\cite{buchman00, saraf16, ziegler14}, and subsequent in-flight performance~\cite{buchman15, saraf16, armano18}. This study reports on a proposed `working' model to describe the sub-threshold electron emission from gold (section~\ref{model}), an experimental measurement campaign to validate the model (section~\ref{expt}) and the results from it (section~\ref{tests}). In section~\ref{discussion} recommendations are made to ensure a robust and predictable solution for a space-based gravitational wave mission, such as LISA~\cite{Amaro17}.    

\section{A Surface Physics Model}
\label{model}
Although the concept of photoelectric emission is relatively straightforward, its application in a space mission is complicated by practical constraints.  For example there are only a limited number of space qualified UV light sources and there is a limit to how far towards short wavelengths that UV fibre-optic cables can be used, if they form part of the photon delivery system, as was indeed the case with GP-B and LISA Pathfinder.  Hence, it is almost inevitable that UV photons with energies below the nominal work function of gold, which is chosen for reasons of stability and inertness, must be used.   However, it is known that surface monolayers of polar molecules, including water~\cite{wells72} and some so-called `adventitious' hydrocarbons~\cite{barr1995, alloway09}  can reduce the effective work function of metals, such as gold.  For water layers on gold the largest effect is from the first chemisorbed monolayer with a less pronounced, but progressive, lowering as more physisorbed layers are added. The physisorbed layers are weakly bonded and evaporate rapidly in vacuum ($\sim\mu$s) leaving only the more resilient chemisorbed monolayer, which can only be fully removed by a high-temperature bake out.  O'Hanlon \cite{hanlon05} suggests that a temperature as high as $\sim250^\circ$C is required, whereas \cite{wells72} noted that most could be removed at $\sim150^\circ$C and derived a desorption energy of $\sim25$\,kcal/mol for freshly deposited water on a clean gold surface cleaved in vacuum.  However, it was subsequently noted that air-contaminated gold surfaces were stable as photoelectron emitters for relatively long periods of time ($\sim 4-5$\,weeks)~\cite{saville95,jiang98} but with a substantial initial decline in the quantum yield over the first $2-3$\,weeks~\cite{jiang98}.  The initial decline was attributed to the possible growth of a contaminant layer in vacuum which could give additional polar species and/or impede the escape of the photoelectrons.  Here we report on measurements done over timescales of months to years which adds hitherto unseen phenomenologies to the behaviour.  However, the basic scenario involving a polar monolayer is retained and the consequences are investigated in terms of design drivers relevant to any long duration practical application of the effect for charge control.  The design drivers are:- \\
1. Quantum yield as a function of photon energy, differentiating between extrinsic and intrinsic quantum yields, $QY_{ext}$ and $QY_{ int}$,  defined as emitted photoelectrons per incident photon and per absorbed photon respectively. \\ 
2. The energy distribution of the photoelectrons, which impacts their transport within electric fields.  \\
3. The dependence of $QY_{ext}$ on angle of incidence.  This must be known in order to model the process as the photons scatter around from surface to surface. \\
4. The dependence of $QY_{ext}$ on surface finish, as surfaces used might be of different quality.  \\
5. The evolution of $QY_{ext}$ and $\phi$ over long periods in vacuum, including the likely effects of prior surface preparations, such as cleaning, storage and bake-out.  

\subsection{Quantum yield}
 Consider a small area element, $dxdy$, of a surface illuminated at normal incidence with photons of energy, $h\nu$. The photocurrent, $dI_{pe}$, assuming a linear density of states below the Fermi energy, is expected to be~\cite{smith74} 
\begin{equation}
\label{dI}
dI_{pe} = kN_{\nu}(x,y)d\nu\left(h\nu - \phi(x,y) \right)^2(1-R(\nu))p_{esc}dxdy
\end{equation}
where $k$ is a constant, $N_{\nu}(x,y)d\nu$ is the incident photon spectral number density, which is taken to have a spatial profile, and work function $\phi(x,y)$ is allowed to be position dependent, $R(\nu)$ is the normal incidence reflectivity, and $p_{esc}$ is the photoelectron escape factor from the surface.  For illumination at an angle, $\alpha$, both the reflectivity and the escape factor will be modified and the equation becomes
\begin{equation}
\label{dIalpha}
dI_{pe} = kN_{\nu}(x,y, \alpha) \left(h\nu - \phi(x,y) \right)^2(1-R(\nu, \alpha))p_{esc}(\alpha)d\nu d\alpha dxdy
\end{equation}

The total photocurrent is then 
\begin{equation}
\label{Ipe}
I_{pe} = \int dI_{pe}
\end{equation}  
where the integration must be done over the whole surface illuminated area, over the whole spectrum of the UV light source, and over the incoming angle of incidence range.   The extrinsic quantum yield is then 
\begin{equation}
\label{QYe}
QY_{ext} = I_{pe}/\int N_{\nu}(x, y, \alpha) d\alpha dxdyd\nu
\end{equation}
Assuming each photon only impacts the surface once, the intrinsic quantum yield is then
\begin{equation}
\label{QYi}
QY_{int} = QY_{ext}/\left(1-R_{average}\right)
\end{equation}
where $R_{average}$ is the reflectivity averaged over the incoming photon energy spectrum, polarisation and angle of incidence.  A typical normal incidence reflectivity from a thin gold film in the UV range studied here, is $\sim 35$\,\%.~\cite{johnson72, zombeck90} There is a slow increase up to $\sim 40$\,\% at $60^{\mbox{o}}$ incidence~\cite{hollington11}. 

A number of compact UV light source technologies are now available for space use.  These include low-pressure Hg discharge lamps as used on ROSAT~\cite{adams87}, GP-B~\cite{buchman95} and LISA Pathfinder~\cite{wass06}, and various deep UV light emitting diodes (LEDs)~\cite{sun06, pollack10, taiwo15, hollington15, hollington17}. Of particular interest is that some of the new LEDs~\cite{taiwo15, hollington15, hollington17} offer spectra with photons of higher energy than the nominal gold work function of 5.1\,eV.  In addition the LED sources emit a broader spectrum than the Hg lamps, which are essentially atomic line sources. Hence equation~\ref{QYe} must be integrated over the source photon energy spectrum.  However, the reflectivity and $p_{esc}$ depend only weakly on photon energy in the range considered here and so 

\begin{equation}
\label{QYeenergy}
QY_{ext} \propto  \left( h\nu - \phi \right)^2
\end{equation}

The angular dependence of the escape factor, $p_{esc}(\alpha)$ is due to two effects.  Increasing the angle decreases the vertical depth penetration of the UV photons into the surface.   These photons have a mean-free path for absorption of $p_{UV}$ which is typically $13-12$\,nm for the photon energies used~\footnote{http://refractiveindex.info}.  The sub-surface electrons released by these photons will then have less distance to travel before escaping. The inelastic mean-free path, IMFP, of these low-energy electrons is typically $\lambda_s = 8-5$\,nm~\cite{Somorjai81}.    The escape factor is then

\begin{equation}
\label{pesc}
p_{esc}(\alpha) = \frac{1}{1+\frac{p_{UV}}{\lambda_s}\cos \alpha}
\end{equation}

$QY_{int}$ increases with increasing incidence angle due to this escape factor.   The ratio of $p_{UV}/\lambda_s$ is of order 1 to 2.
 At our largest reference angle of $60^{\mbox{o}}$ the enhancement has a maximum value of 1.6 and 95\% of the photoelectrons originate from within 10\,nm of the surface.

\subsection{Energy distribution of emitted electrons}
For a given energy photon, the energy spectrum of emitted electrons will, in principle, start at zero energy and extend up to a maximum determined by the difference between the photon energy and the work function.  The basic shape of the energy distribution will be governed by the densities of state within the occupied valence band, from which electrons are extracted and the largely unoccupied conduction band into which the electrons are lifted.  This basic starting shape will be modified away from the ideal zero temperature distribution by thermal broadening ($\sim$20\,meV at room temperature). Once liberated, the electrons then need to migrate to the surface and this process is characterised by the inelastic scattering mean free path (IMFP), $\lambda_s$, \cite{Somorjai81} which will degrade the emitted electron energy spectrum. On reaching the surface the electron escape probability will be dependent on the electron energy including possible quantum tunnelling effects \cite{hechenblaikner12}. Finally, any finite spectral line width from the UV illumination source will cause additional broadening.  These effects will be discussed further when the results of the test campaign are presented in section~\ref{tests}.

\subsection{Angular distribution of emitted electrons}
Since the photon energy is comparable to the work function of the gold surface, the angular distribution of emitted electrons can be described by a cosine function, as shown by Pei and Berglund~\cite{pei02}.
\begin{equation}
\label{eangle}
I_e\left(\beta\right) = I_{e0}\cos \beta
\end{equation}
Here $\beta$  is the emission angle relative to the normal.  Knowing this angular distribution can be important in assessing the effectiveness of applied electric fields to control the electron transport in some applications~\cite{hollington11, ziegler14}.

\subsection{Work function variation}
Equation~\ref{QYeenergy} assumes a surface with a uniform work function over the whole of the illuminated region. If instead the surface had a patchy covering of adsorbents, $i$, then the intrinsic quantum yield would be 
\begin{equation}
\label{qypatch}
QY_{ext} \propto A_g\left( h\nu - \phi_g \right)^2 + \Sigma_i A_i\left( h\nu - \phi_i \right)^2
\end{equation}
where $A_i$ is the area affected by adsorbent $i$ and $\phi_i$ is the work function for that particular species.  The first term allows for the presence of uncontaminated gold. \\
One of the original motivations for this study was to test this possibility, as an argument for using very short wavelength UV to mitigate the uncertainties from poorly controlled surface adsorbents by making the first term dominant.

\section{\label{expt}Experimental Method}
This work was part of a wider investigation, initially funded by the European Space Agency (ESA),  into charge management for its L3 gravitational wave mission.  Key elements were studying new UV light sources~\cite{hollington15, hollington17}, numerical modelling and carring out representative discharge measurements.  The numerical modelling was to follow UV photons as they scatter around inside a complex geometry, photoelectron production from a number of segmented surfaces, and tracking the subsequent drift of the electrons within applied ac electric fields present.  Modelling of this sort (\cite{hollington11, ziegler14}) must incorporate accurate geometrical and physical models for all processes, including photoelectric emission process from realistic large area surfaces, which need to be `handled' outside of vacuum at various stages and can be made with reliable fabrication procedures.   Surfaces used here were of similar specification to those in LISA Pathfinder and measurements were done in a dedicated vacuum test rig with calibrated UV light sources with wavelengths between 200\,nm and 270\,nm.  Measurements included photoelectic yield, surface work function determination, electron energy distribution, angular dependence and surface roughness effects.  

\subsection{Test Surfaces} 

A set of three 46\,mm square surfaces were defined. Copper substrates, 5\,mm thick, were procured commercially~\footnote{Mateck: http://www.mateck.com/} to three surface roughness specifications typical of those present in a real flight instrument. After supply they were inspected and characterized for surface finish.  Gold coating, to the LISAPF specification, was applied commercially~\footnote{Teer Coatings:  http://www.teercoatings.co.uk/} and the surfaces were kept in clean, sealed conditions until use.  Table~\ref{surfaces} shows the specifications and the measured roughness parameter, $R_a$, over the area to be illuminated by the UV.  Surface roughness was measured prior to coating to minimise handling of the coated samples.  Connections for the surfaces were via two metal contact pins located into the rear surface. 
\begin{table}[h]
\caption{\label{surfaces}Roughness of gold plated test surfaces used }
\begin{ruledtabular}
\begin{tabular}{ccc}
Surface  & Specification & Roughness \\ 
& (nm) & (nm) \\ \hline
MTK010 & $<$30 & 7~\footnote{Using a white light ZYGO interferometer} \\
MTK200 & 250 & 130~\footnote{\label{talysurf}Using Talysurf profilometer measurements} \\
MTK800 & 800 & 340~\footnotemark[2] \\
\end{tabular}
\end{ruledtabular}
\end{table}

\begin{figure}[h]
\includegraphics[width=3.4in]{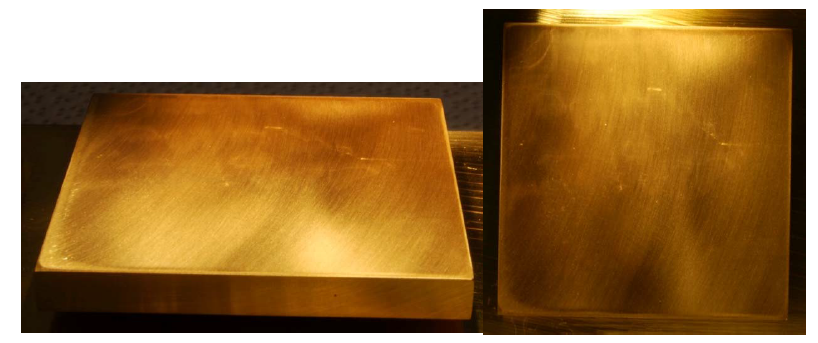}%
\caption{\label{surfacephoto}Photographs of MTK200-01 illuminated at normal (left) and grazing incidence (right).}
\end{figure}

Prior to coating substrates were cleaned in an ultrasound bath using Citranox~\footnote{https://www.alconox.com}, rinsed in water, with a second ultrasound in warm ethanol, after which they were held vertically whilst the ethanol evaporated leaving a stain and residue-free surface. After cleaning the substrates were mounted into individual sealable aluminium containers, and inspected under black-light for particle contamination. The containers were then closed, flushed with dry nitrogen gas and sealed. At the coating supplier 50\,nm titanium and 800\,nm of gold were applied to the each substrate as well as onto a quartz witness plate.   The surfaces were returned to the transport containers which were sealed in evacuated bags.  On receipt the first samples were photographed under strong white light. Figure~\ref{surfacephoto} shows the mid-roughness sample, MTK200-01, where some residual machining markings on the substrate can be seen.  Any particle contamination was easily removed at first inspection.   

\subsection{Vacuum Test Rig}  
\begin{figure}[h]
\includegraphics[width=2.6in]{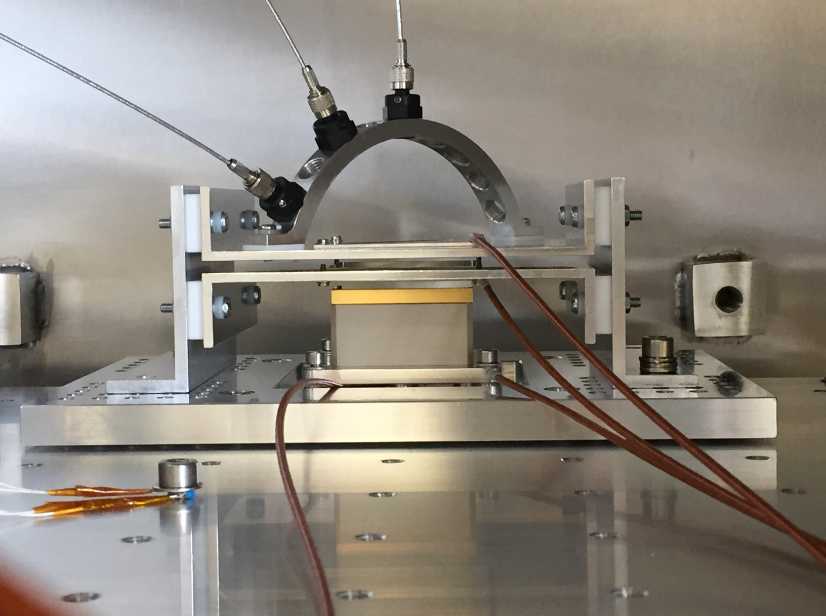}
\includegraphics[width=3.8in]{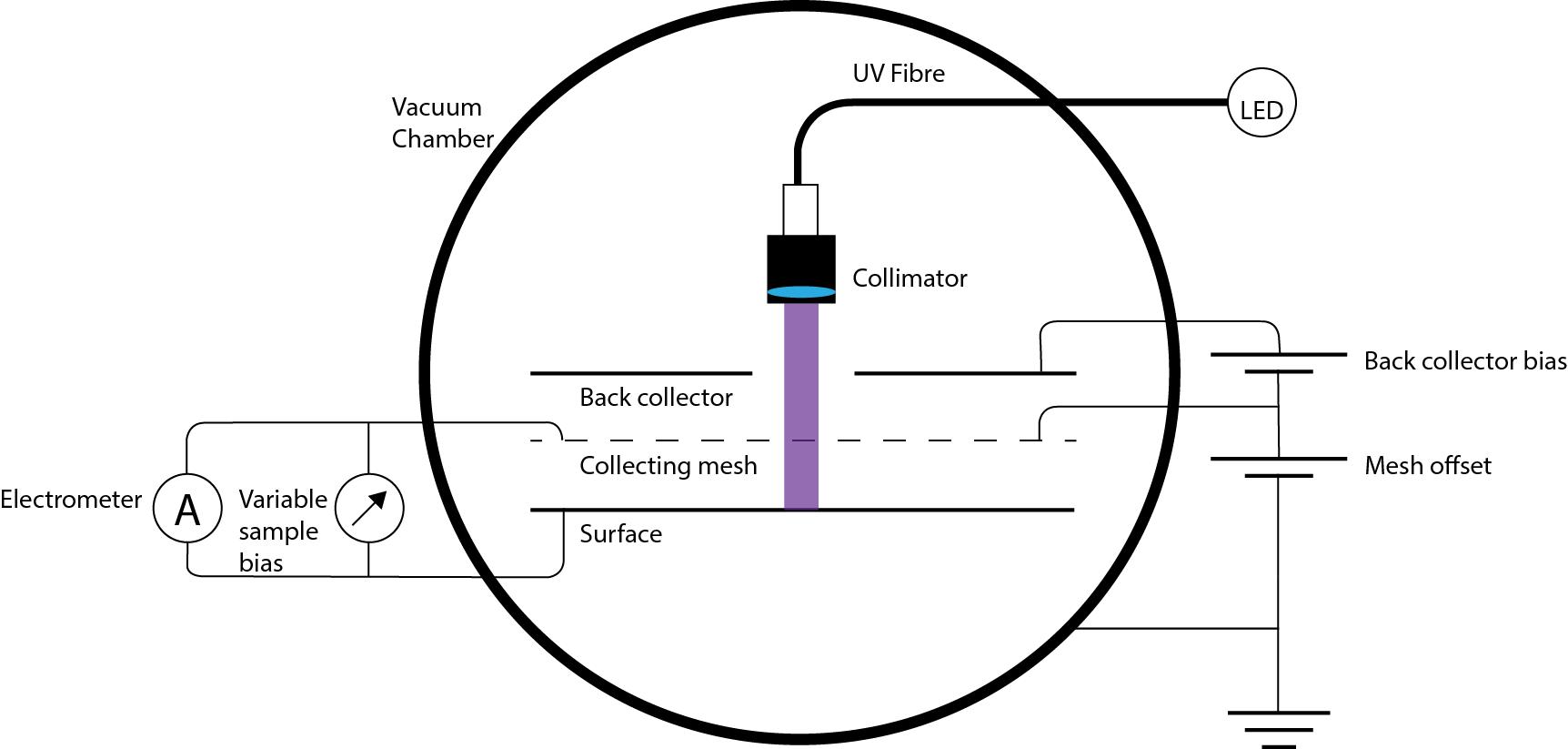}%
\caption{\label{qyvacsys}  Left: The measurement jig with three fibre angles populated. Right: Arrangement of the photocurrent measurement grids.}
\end{figure}

A jig, which could hold a surface with fibre optic cable feeds to provide illumination at a range of fixed angles as shown in figure~\ref{qyvacsys} (left), was installed into an oil-free vacuum chamber with a base-vacuum of $<10^{-7}$\,mbar. The fibre optic cables were brought out through fixed vacuum interfaces and a range of external UV light sources could be used.  \\

A collecting electrode above the surface with programmable bias allowed the photoelectron energy distribution and $QY_{ext}$ to be obtained from measurement of the total drain current from the surface.  UV light was fed from the source to the internal measurement setup (shown in Figure~\ref{qyvacsys} (right)) by UV fibre optics. The light was injected onto the surfaces using a collimating optic (Avantes COL UV/VIS).  The setup allowed measurements at three angles of incidence ($0^\circ$, $30^\circ$, $60^\circ$) which were preset prior to closing the vacuum system.  The current between surface and the collecting grid was measured using an electrometer (Keithley 6514) with a pulsed UV source and phase sensitive demodulation to enhance the sensitivity.  A second capture grid was used to trap electrons emitted from the collection grid.  Before any surfaces were installed in the chamber the system was tested without surfaces to establish the baseline sensitivity of the technique.  Using a modulated (on/off) UV light source with 200\,s period and 50\% duty cycle, a 12\,hour measurement achieved a sensitivity of 0.05\,fA.


\subsection{Calibrated UV light sources}
Table~\ref{sources} gives details of the UV light sources used which included three UV LEDs, a low-pressure Hg discharge lamp of the type used on LISAPF, and a broadband deuterium lamp together with a UV Spectrometer/Monochromator.

\begin{table}[h]%
\caption{\label{sources}Wavelengths used during quantum yield measurements}
\begin{ruledtabular}
\begin{tabular}{cccc}
Device  & Manufacturer & Peak wavelength & FWHM \\
& & (nm)  & (nm) \\ \hline
SET240 & SET Inc\footnote{Sensor Electronic Technology Incorporated, Deep UV LED Catalogue, Data sheet, 
	http://www.s-et.com/uvtop-catalogue.pdf 2011
} & 247.0 & 10.5 \\
CIS250 & Crystal IS\footnote{Crystal IS, Optan TO-39 UVC LEDs, Data sheet,   http://www.cisuvc.com/, 2014} & 249.8 &11.5 \\
Hg lamp & UV Products,\footnote{http://www.uvp.com/penraylightsources.html} Penray & 253.7 & $<1$ \\
SET255 & SET Inc & 258.5 & 11.0 \\
$^2$H lamp +  & Newport 63165 & 200 to 270 &$\sim$5 \\
monchromator &  Princeton Instr.\footnote{Customised in-house}       &     &  \\
\end{tabular}
\end{ruledtabular}
\end{table}

\subsection{Sequence of Measurements}
Measurements were split into two phases with a long gap between them.  During Phase 1 the surfaces were characterised for their detailed behaviour under different UV illumination schemes.  During Phase 2 some longer term stability studies were made.

\subsubsection {Phase 1}  
Phase 1 was focused  on checking the basic properties of the physical model.  The first sample was MKT200-01 and this was installed and the tests shown in Table~\ref{qytests} were carried out in the sequence shown over a 4-week period.  Not shown for clarity were Hg lamp reference  measurements at $\alpha =0^\circ$ done repeatedly throughout the period.  All measurements included a full scan of bias voltages.
Once these were completed the chamber was opened and the second surface installed (MKT010-01).  A slightly extended sequence of measurements was done over the same time period.  Finally the third surface, MKT800-01, was installed and measured. The yield was seen to be time-dependent for all three surfaces and hence the sequencing and time locations were kept as similar as possible to allow useful comparisons. Having said that, some building works outside of our control did impact on the final surface campaign timing.  Although this was undesirable it did actually provide some useful diagnostic information.  After the nominal phase 1 campaign with MTK800-01 was completed the surface was monitored over a long extended time period, for the first time for this type of measurement, to look for longer term changes. A liquid nitrogen cold plate was used towards the end of that period to look for any effect following a reduction in the partial pressure of water vapour in the system.  Subsequent to these measurements a second period of investigations was undertaken to follow-up on the long-term evolution of the quantum yield and work function results in vacuum.
\begin{table}[h]
\caption{\label{qytests}QY Test campaign for the three surfaces.}
\begin{ruledtabular}
\begin{tabular}{cc}
Light Sources & Angles of incidence  \\ \hline
SET240 & $0^{\mbox{o}}$, $30^{\mbox{o}}$, $60^{\mbox{o}}$     \\
CIS250 & $0^{\mbox{o}}$, $30^{\mbox{o}}$, $60^{\mbox{o}}$     \\
SET255 & $0^{\mbox{o}}$, $30^{\mbox{o}}$, $60^{\mbox{o}}$     \\
Hg lamp & $0^{\mbox{o}}$, $30^{\mbox{o}}$, $60^{\mbox{o}}$     \\
Hg lamp, SET240, CIS250, SET255, $^2$H lamp + monochromator  & $0^{\mbox{o}}$ \\
\end{tabular}
\end{ruledtabular}
\end{table}

The sequence of tests outlined in table~\ref{qytests} were designed to 
a) measure the quantum yield of the surfaces using the different UVLED devices, 
b) probe the photoelectron energy distribution by varying the bias potentials,
c) test the dependence on the angle of incidence,
d) test the effect of surface roughness, and 
e) test the effect of wavelength on quantum yield.

\subsubsection{Phase 2}
During phase 2 the quantum yield and work function were measured over several weeks for a number of surfaces to further investigate the time dependent behaviours seen in phase 1.  These measurements were done on the same surfaces from phase 1, but with up to 3 years time interval, and also on some unused surfaces produced as part of the initial batch.  In addition the effects of some surface preparations and in vacuum bake-out procedures were investigated.  In total eight long duration investigations of the time evolution were carried out as detailed in table~\ref{phase2runs}.  The first three of these were during phase 1. The first measurement in phase 2 showed no detectable emission and was terminated after just 4\,days.  All measurements in phase 2 were done using UV illumination from the CIS250 LED.

\begin{table}[h]
\caption{\label{phase2runs}Investigations of the time evolution of yield and work function, in chronological order}
\begin{ruledtabular}
\begin{tabular}{cccc}
Run & Surface  & Duration & Comment  \\ 
& & (days) &  \\ \hline
1 & MTK200-01 & $28$ & Part of phase 1 campaign \\
2 & MTK010-01 & $28$ & Part of phase 1 campaign \\
3 & MTK800-01 & $35$ & Part of phase 1 campaign. With unplanned vacuum outages.  \\
& & &  Use of liquid nitrogen cold trap \\ \hline
4~\footnote{After 27 months in storage} & MTK800-01 & $4$ & No detectable yield\\ 
5 & MTK010-01 & $63$ & Including the effect of a short air exposure at the end\\
6 & MTK800-01 & $60$ & Including four bake-out cycles at $115^{\mbox{o}}$\,C\\
7 & MTK010-02 & $66$ & Including study of effects of pressure excursions\\
8 & MTK010-02 & $40$ & With prolonged bake-out between $150^{\mbox{o}}$\,C and $200^{\mbox{o}}$\,C\\
9 & MTK010-03 & $22$ & With prolonged bake-out between $150^{\mbox{o}}$\,C and $200^{\mbox{o}}$\,C\\

\end{tabular}
\end{ruledtabular}
\end{table}

\section{\label{tests}Test Campaign Measurements}  

\begin{figure}[h]
\includegraphics[width=6.2in]{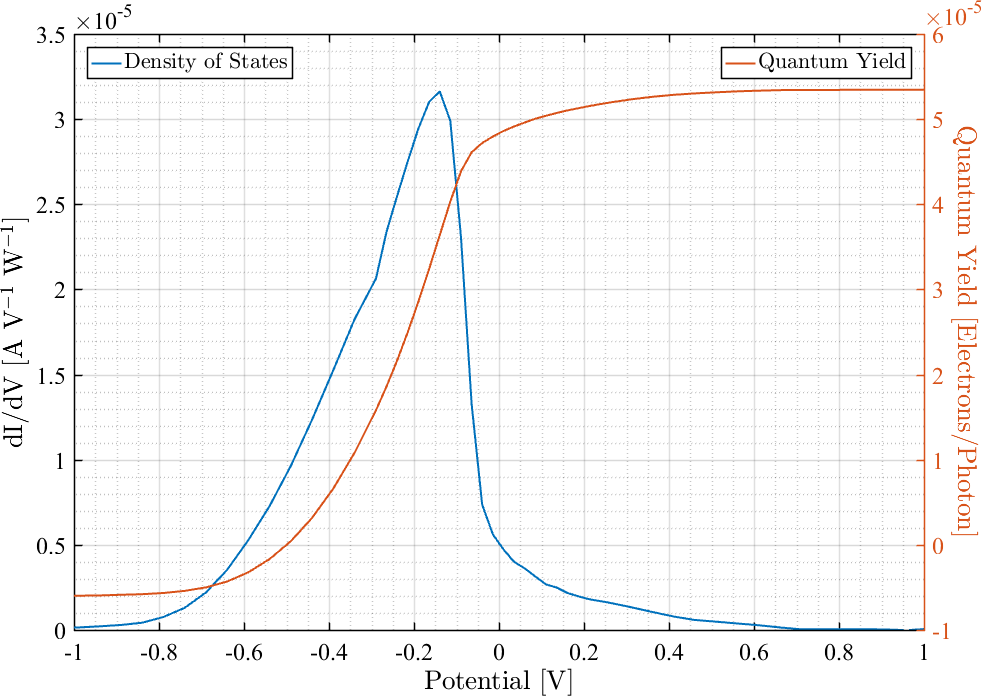}%
\caption{\label{voltscan}  A plot of yield and density of states taken from a collector voltage scan using the SET240 device at 30$^\circ$ incidence.}
\end{figure}

 An example of a typical collector grid bias voltage scan is shown in figure~\ref{voltscan}. The bias voltage is applied to the collector grid.   The photoemission current, expressed as a quantum yield, is the red curve, and its differential is the blue curve. 
At -1\,V bias most of the photoelectrons being emitted are being repelled back.  The small residual photocurrent is due to reflected UV reaching the collector grid itself giving a reverse photocurrent.  As the bias is made more positive the photocurrent from the surface begins to increase and reaches a plateau by the time the bias is at +1\,V, at which value all of the emitted photoelectrons are being measured.  This is referred to as the saturated photocurrent and gives a measure of $QY_{ext}$.   The differential with respect to bias voltage is also shown in the figure and the detailed shape of this curve gives a measurement of the emitted electron energy distribution, which gives a measure of the effective work function.  

\subsection{Phase 1 - Physics Model Validation}
\begin{figure}[h]
\includegraphics[width=3.2in]{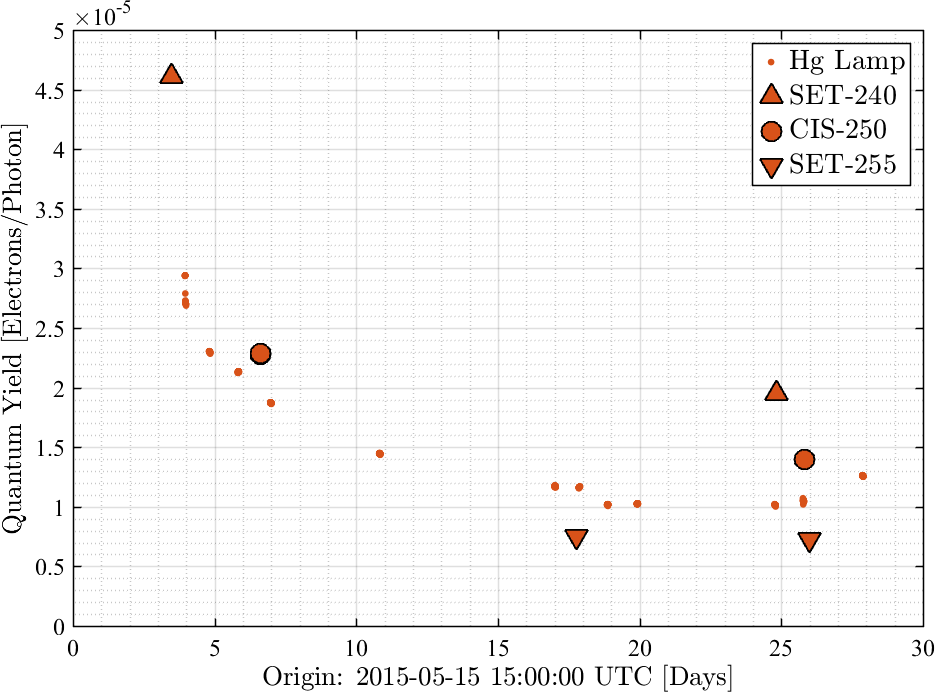}
\includegraphics[width=3.2in]{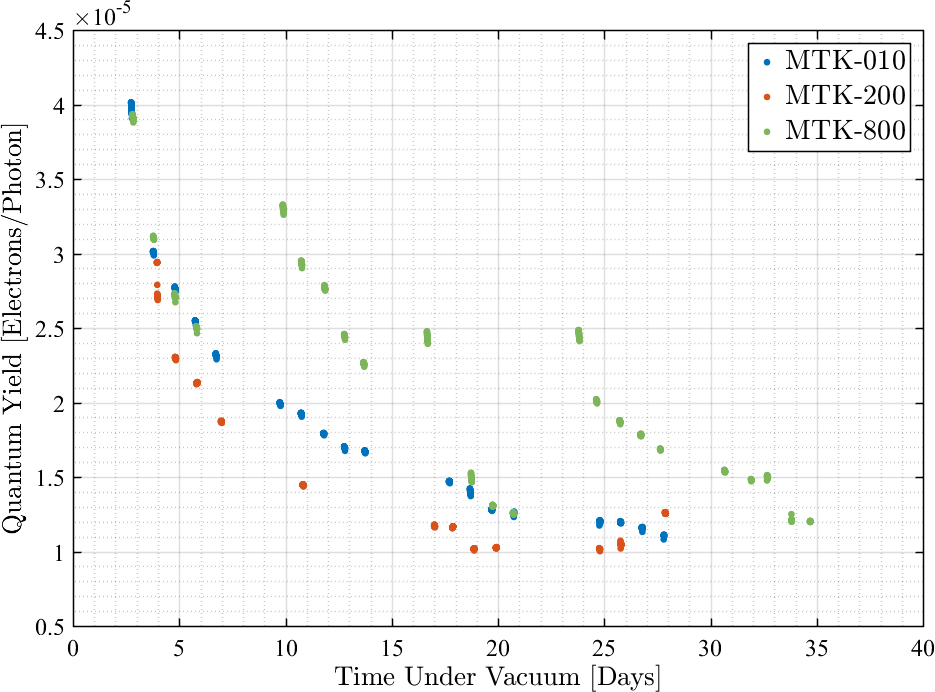}
\caption{\label{yields} (Left) Saturation $QY_{ext}$ measurements from the 4 week study period for MTK200-01. (Right) Saturation $QY_{ext}$ measurements at normal incidence from MTK010-01, MTK200-01 and MTK800-01 using a Hg lamp.}
\end{figure}

Figure~\ref{yields} (left) shows the time sequencing of saturated $QY_{ext}$ measurements for the first surface tested, MTK200-01.  Most measurements were done using the Hg lamp as a reference.  The figure shows two sets of data taken with the three LED types.  

The time decay in the results was the first time this had been explicitly observed in such a clear way, and had two consequences for the subsequent phase 1 test campaign.  Firstly, it complicated the comparative data analysis and for this reason the campaign was extended until a stable behaviour was seen, after which the final definitive measurements were done.  Secondly, the other two phase 1 surface campaigns were then made as similar as possible given other logistical constraints.  

\subsubsection{Comparison reference measurements using the Hg lamp}
Figure~\ref{yields} (right) shows the saturated $QY_{ext}$  data using the Hg lamp $\left( 2537\,\mbox{\AA} \right)$ at normal incidence on all three surfaces overlaid with a common start time to allow comparison of the behaviour. It can be seen that the three surfaces behaved similarly, with starting $QY_{ext}$ values between 40 and 55\,ppm (photoelectrons/incident photon). $QY_{ext}$ reached stability on similar time frames ($\sim$17\,days) with very similar equilibrium values of 10 to 15\,ppm.  \\
The decay could, in principle, be consistent with an effective work function increase as physisorbed polar layers (e.g. water) are outgassed from the surface~\cite{wells72}, but the time constant is too long.~\cite{hanlon05} The change in work function needed would be $\sim0.2$\,eV ($4\%$) which would have been measurable, but which was not seen (see Section~\ref{stabilitysection}), although others have in rather different circumstances~\cite{jiang98}. A similar decay behaviour is seen for MTK800-01 at times of $0.8\times 10^6$\, and $\sim2 \times 10^6$\,s when the vacuum integrity had been compromised by unplanned power outages resulting in loss of vacuum.  Section~\ref{stabilitysection} returns to this discussion.

\subsubsection{Photoelectron energy distribution and work function determination}
The detailed shape of the differential blue curve in Figure~\ref{voltscan} depends on several effects:- \\
1. $(h\nu_m - \phi)$ where $\nu_m$ is the mean frequency of light from the source \\
2. The density of states at the top and bottom of the valence and conduction bands \\
3. Temperature through thermal broadening \\
4. The spectral width of the light source used \\
5. Quantum tunnelling effects   

\begin{figure}[h]
\includegraphics[width=3.2in]{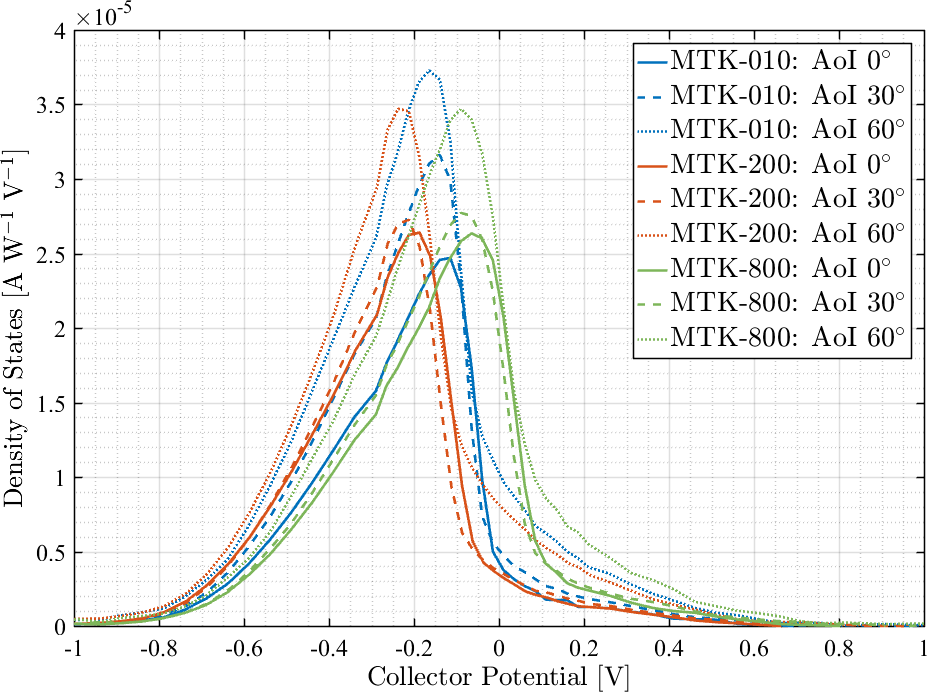}
\includegraphics[width=3.2in]{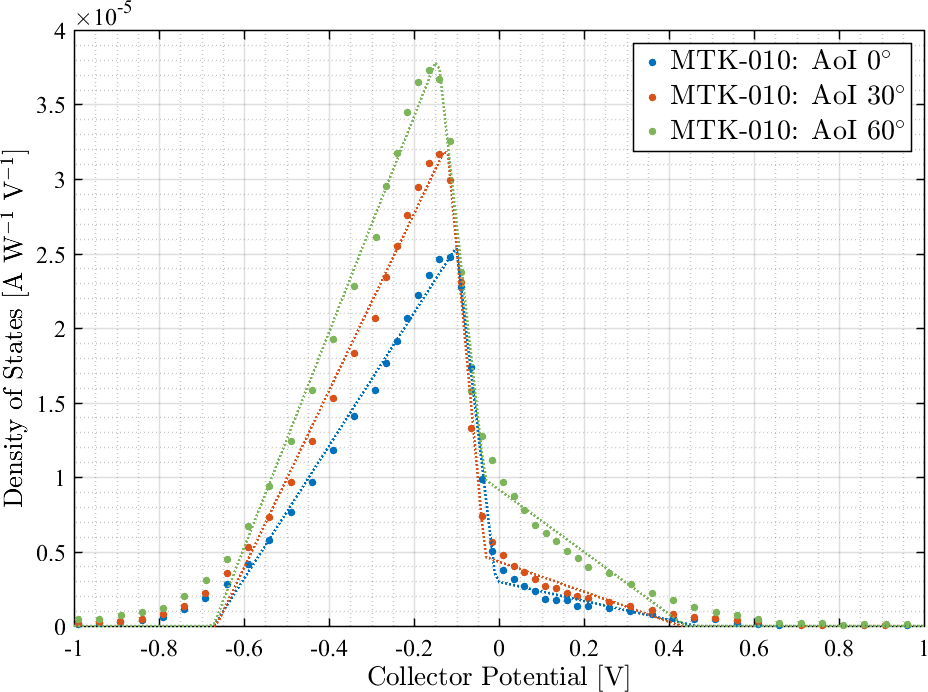}%
\caption{\label{statedens}  Left: Differential scans from all surfaces at three angles of incidence using SET240 illumination. Right: Fits to the SET240 data from MTK010-01 using simple triangular functions.}
\end{figure}

Figure~\ref{statedens} (left) shows plots of the differential curve behaviours using SET240 at three angles of incidence on all three surfaces. Similar data were collected using the other UV sources, CIS250, SET255 and the Hg lamp. Certain general properties of the distribution are notable:-  \\  
1. The peak of the distribution shifts to higher amplitude but lower voltage (higher electron energy) with increasing angle of incidence. \\
2. The width of the distribution does not change obviously at different angles of incidence. \\
3. The relative shapes of the curves at different angles of incidence are comparable across the three different surfaces. \\
4. The position of the distribution shifts between surfaces, moving to higher collector potentials over the course of the measurement campaign. 

A simplified fitting function was applied to the distributions shown in Figure~\ref{statedens} (left), and those from the other light sources, to recover the effective work function for each surface.  Fits to the SET240 data for MTK010-01 are shown in Figure~\ref{statedens} (right) with a simple triangular form, but  including a reverse current component for reflected light releasing photoelectrons from the platinum grid.  

Fits have been done for all measurements taken during phase 1, and the mean effective work function and its standard deviation are shown in Table~\ref{workfs} for each surface. 

\begin{table}[h]
\caption{\label{workfs}The average and standard deviation from all work function measurements made during the initial test  campaign on the first three surfaces}
\begin{ruledtabular}
\begin{tabular}{cccc}
 & MTK200-01 & MTK010-01 & MTK800-01 \\ \hline
$\phi$(eV) & 4.41 & 4.40 & 4.29 \\
$\sigma$(eV) & 0.04 & 0.03 & 0.03 \\
\end{tabular}
\end{ruledtabular}
\end{table}

An effective work function of ~4.4\,eV is not dissimilar to results from many previous measurements on a variety of gold surfaces done within the LISA Pathfinder campaign and had been tentatively ascribed to water on the surfaces reducing the work function.~\cite{jiang98,ziegler13}

\subsubsection{Effect of surface roughness} 
\begin{figure}[h]
\includegraphics[width=3.2in]{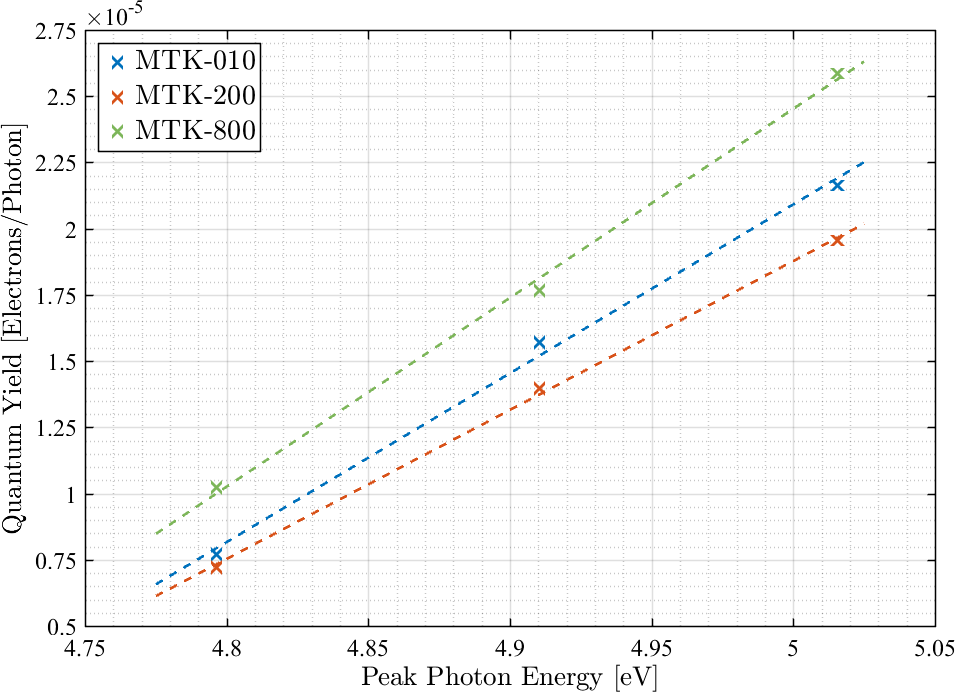}
\includegraphics[width=3.2in]{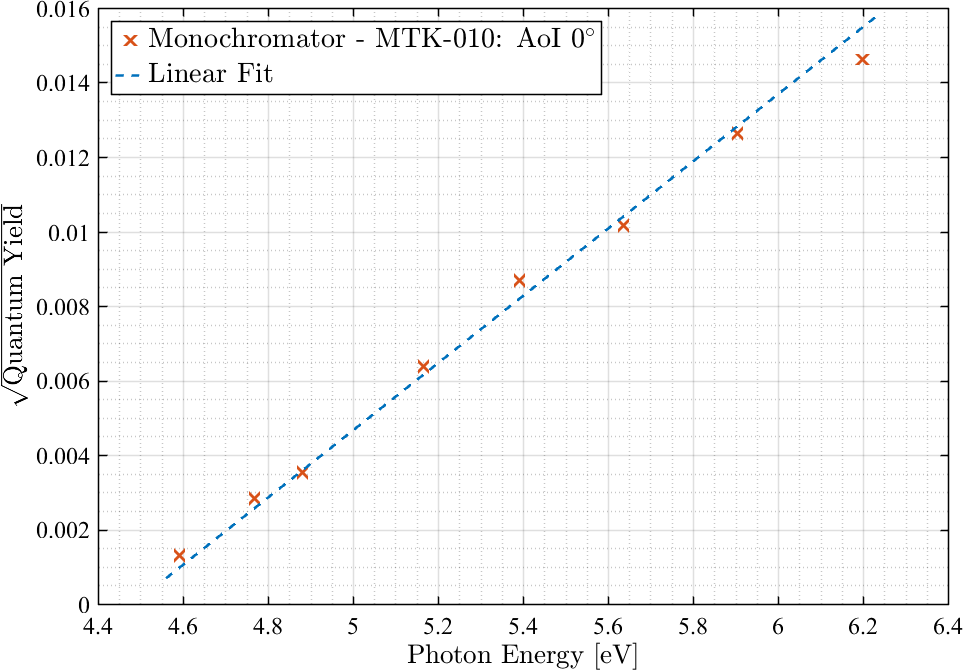}
\caption{\label{variation} (Left)  Normal incidence $QY_{ext}$ variation as a function of roughness for the three LED sources. (Right)  $QY_{ext}^{1/2}$ as a function of photon energy.  The fit assumes a single species with $QY_{ext}^{1/2}\propto$\,$h\nu$.}
\end{figure}

The effect of surface roughness is not significant as can be seen from the normal incidence data using the LED UV sources from the three surfaces of different roughness shown in Figure~\ref{variation} (left). Although there is a difference between the surfaces it is not clearly ordered by roughness.  Compilation of all data at different angles of incidence and using different wavelengths shows a dispersion of $<$10\% confirming that surface roughness is not a significant parameter in terms of photoelectric yield. Theoretically a small effect is predicted whereby multiple scattering can increase the extrinsic yield from very rough surfaces and Figure~\ref{variation} (left) seems to show a small effect within the normal incidence data. 
 However, the smallness of the effect does not mean it is inconsequential for discharging itself as the angular distribution of photon scattering still plays a role.  

\subsubsection{Variation as a function of wavelength}

To obtain a wider range of wavelengths with narrower linewidths than could be obtained from the UV LEDs this measurement was performed  with a deuterium lamp and monochromator on MTK010-01.  The wavelength range covered was extended in both directions to provide more leverage on the physics and to explore the possible benefits of further development of laser diodes to shorter wavelengths. 
The results are shown in Figure~\ref{variation} (right).  The shorter wavelength data have been corrected for fibre absorption.   Data from the UV LEDs are fully consistent with Figure~\ref{variation} (right) once the spectral outputs have been properly integrated over their extended emission ranges which lie between 4.8 and 5.1\,eV

The following observations are made:
\begin{itemize}
\item	A smoothly increasing function with a $QY_{ext}$ of 225\,ppm at 6.2\,eV. 
\item	There is no obvious discontinuity at the nominal gold work-function energy (5.1\,eV).  This argues against any exposed gold surface, providing a new significant source of photoelectrons.
\item	The yield goes to zero at $\sim$4.5\,eV
\item	The fit gives a more accurate determination of the work function of MTK010-01. 
\item  There is no evidence for significant patchiness in the work-function (which would broaden the measured distribution of the density of states).
\end{itemize}
UV LED data were taken from all surfaces allowing a comparison of the work function for each surface using this method.  The combined results are contained in Table~\ref{workf}.  The work functions of all three samples are the same to within the measurement uncertainties and consistent with values obtained from the electron energy distribution measurements shown in table~\ref{workfs} apart from a small difference for MTK800-01.
\begin{table}[h]
\caption{\label{workf}Calculated effective work functions based on the saturated quantum yields}
\begin{ruledtabular}
\begin{tabular}{cccc}
Surface & source & work function & uncertainty \\ \hline
MTK010-01 &  monochromator & 4.43\,eV & $\pm$0.01 \\
MTK010-01 &  UV LED & 4.44\,eV & $\pm$0.06 \\
MTK200-01 & UV LED & 4.42\,eV & $\pm$0.10 \\
MTK800-01 & UV-LED & 4.44\,eV & $\pm$0.23 \\
\end{tabular}
\end{ruledtabular}
\end{table}

\subsubsection{Variation as a function of angle of incidence}
\begin{figure}[h]
\includegraphics[width=6.2in]{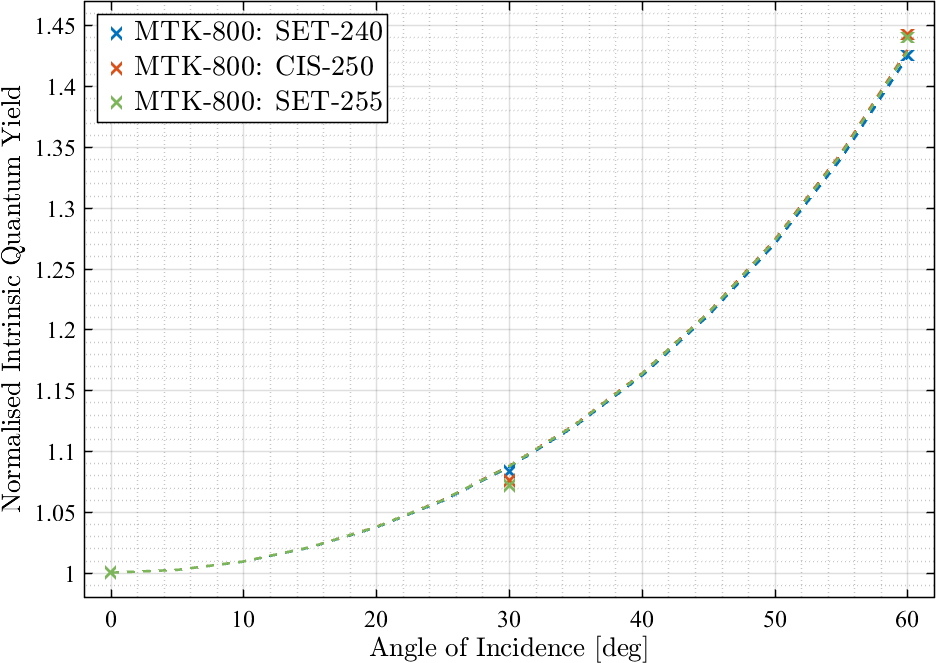}%
\caption{\label{angle} $QY_{int}$ as a function of angle of incidence for MTK800-01.   The data have been corrected for an angular dependence in the reference photodiode readout and surface reflectivity. The dashed curve is a fit of the form given by equation~\ref{pesc}}
\end{figure}

The enhancement factor in $QY_{int}$ due to the angle of incidence for MTK800-01 is shown in Figure~\ref{angle} for all 3 UV LED wavelengths. These data have been corrected for the reflectivity and reference photodiode response.  $QY_{int}$ is seen to increase in going to higher angles of incidence.  This is true for all 9 data sets (3 wavelengths for 3 surface roughnesses).  
 
Using Figure~\ref{angle}, the functional form of the increase from 0$^\circ$ to 60$^\circ$ is well determined and a compilation of values from all surfaces is given in Table~\ref{angletab}. 
The increase is a combination of greater reflectivity (lowering $QY_{ext}$) and shallower UV penetration with easier escape of the photoelectrons  
(increasing $QY_{int}$ and $QY_{ext}$) as described earlier.   

The mean enhancement factor at $60^{\mbox{o}}$ is compatible with the largest values expected from equation~\ref{pesc} but does not have enough fidelity to define the ratio $p/\lambda_s$ to better than between 1 and 4.
\begin{table}[h]
\caption{\label{angletab}Angular dependence of $QY_{int}$ between 0 and $60^{\mbox{o}}$}
\begin{ruledtabular}
\begin{tabular}{ccc}
UVLED & Surface & Relative increase from 0 to 60$^\circ$\\ \hline
SET240 & MTK010-01 & 1.5 \\
CIS250 & MTK010-01 & 1.2 \\
SET255 & MTK010-01 & 1.5 \\
SET240 & MTK200-01 & 1.6 \\
CIS250 & MTK200-01 & 1.5 \\
SET255 & MTK200-01 & 1.7 \\
SET240 & MTK800-01 & 1.4 \\
CIS250 & MTK800-01 & 1.4 \\
SET255 & MTK800-01 & 1.4 \\
\end{tabular}
\end{ruledtabular}
\end{table}

\subsection{Phase 2 - Long Term Variations}
\label{stabilitysection}

\subsubsection{Quantum Yield and Work Function Variations}

Figure~\ref{yields} (right) shows the first look at stability in $QY_{ext}$ obtained during phase 1.  The starting yields, reduction factor and decay time constants are not dissimilar to those seen previously.~\cite{jiang98}  The overall time coverage of the measurements is also similar.  However, the observation of the behaviour of MTK800-01 on accidental multiple short exposures to air was unexpected and not consistent with prior suggestions that additional contamination was likely responsible for the reductions in yield happening after the samples are placed in vacuum.  The data from this work show that the yield is restored on re-exposure to air very quickly and that the subsequent reduction starts anew with approximately the same time constant.  The time constant is roughly 7\,days implying an activation energy $\sim1.1$\,eV for any thermally induced relaxation.  This is not dissimilar to the assumed desorption energy of the last water monolayer~\cite{wells72} but loss of that should continue until there was no further yield at all from the UV sources with photon energies below the nominal gold work function.  During phase 1, work function measurements were not made frequently enough to track its evolution through the initial decay period or with sufficient fidelity to see the required $\sim0.2$\,eV increase in its value to explain the variation.  The measurements quoted earlier for the work function derive mainly from measurements done after the yield had stabilised and these are similar to other results~\cite{wells72, saville95, jiang98, hechenblaikner12}.    

Following phase 1, funded by ESA, there was a long break, due to intense activity with LISA Pathfinder ground activities, launch and mission operations.  During that period MTK800-01, had remained in the vacuum chamber, but not in a controlled vacuum.  

It was some 27 months later that another $QY_{ext}$ measurement was done on MTK800-01 using CIS250 at normal incidence.  The photocurrent was unmeasurable ($< 0.05$\,fA) and remained so for a number of days in vacuum (Run 4 in table~\ref{phase2runs}) .  

\begin{figure}[h]
\includegraphics[width=6.2in]{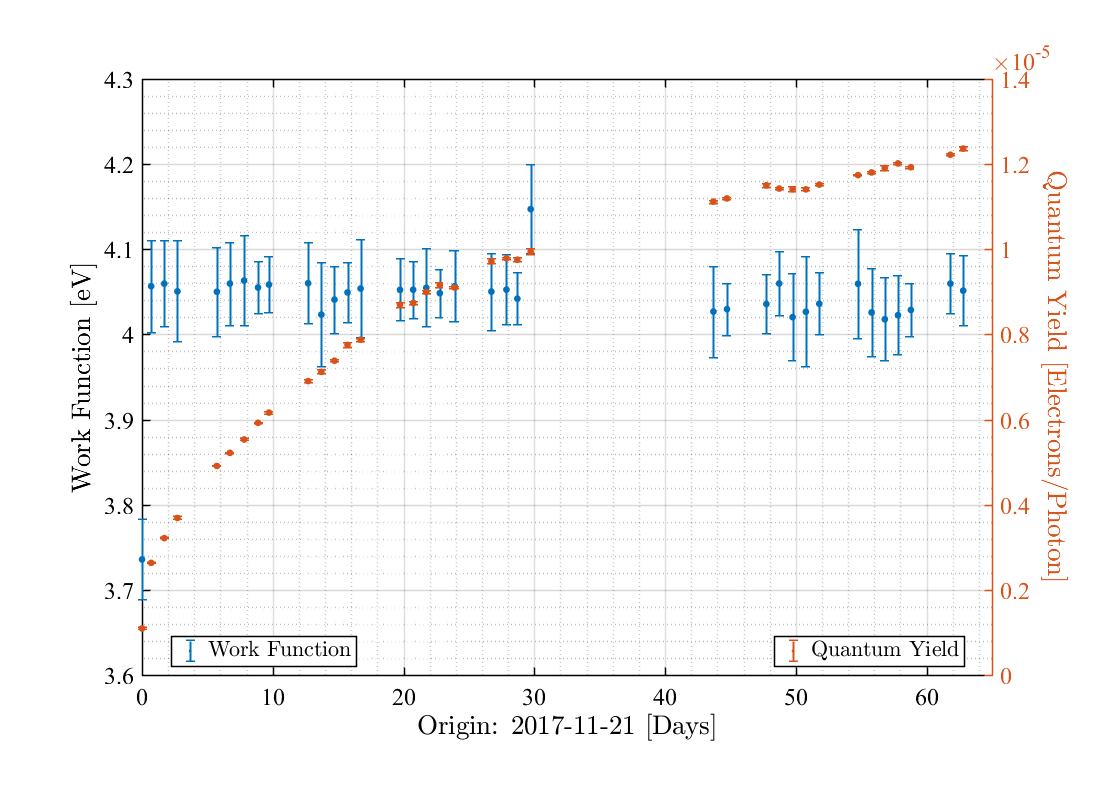}
\caption{\label{MTK010yield2}  Quantum yield from MTK010-01 using CIS250 after 27 months in storage but with a second clean.  The plot shows a 63 day period of repeated $QY_{ext}$ and work function measurements.  }
\end{figure}

Surface MTK010-01 was then re-cleaned and measured over a period of 10 weeks (Run 5).  The result is shown in figure~\ref{MTK010yield2}. Once again there is a long time constant before a stable state is reached, but this time the yield increases from a low value.  Recalling figure~\ref{yields} (right) it can be seen that the time constant is longer ($\sim17$\,days) but that the asymptotic value is similar even though it is approached from below.  This behaviour is clearly no longer consistent with the notion of vacuum evaporation of either physisorbed or chemisorbed water layers causing the initial change in yield.  Neither is the fact that the work function measurements show very little change whilst the quantum yield changes by a factor of 5. Such a large increase would have required a $\sim$\,70\% reduction in work function, which is not seen. It is also evident that the quantum yield increase is not caused by the UV photons themselves as the increase is continuing through periods (weekends) when no measurements were done.  The work function measured is $\sim4.1$\,eV which is significantly different from those in table~\ref{workfs} suggesting that the enabling (polar) surface adsorbent is not water in this case.  The cleaning procedure used was the same as that used before the initial measurements on the same surface leading to the results of figure~\ref{yields} (right) and table~\ref{workfs}.  The difference being that this surface had been exposed to an air environment for some 27 months between the measurements, giving the opportunity for new surface adsorbents. The surface was then removed from the chamber and put into clean storage in air for several days.  Measurements made over the next two days (not shown on the plot) 
showed a dramatic initial reduction in $QY_{ext}$ but with a faster recovery time constant than before.  This is reminiscent of the unplanned observation with MTK800-01, seen in figure~\ref{yields} (right) when, on two occasions, the vacuum integrity was compromised.  Both times the immediate effect was to instantaneously {\it increase} the quantum yield back towards its starting value from where the long time-constant decline continued once the vacuum was re-established. The fact that the effect seen for both surfaces was to reset back towards the initial starting quantum yield value, which was high in one case and low in the other suggests that this is not immediately associated with the air-borne concentrations of the particular active adsorbents but may be more to do with disturbing equilibrium distributions in vacuum.

\begin{figure}[h]
\includegraphics[width=6.2in]{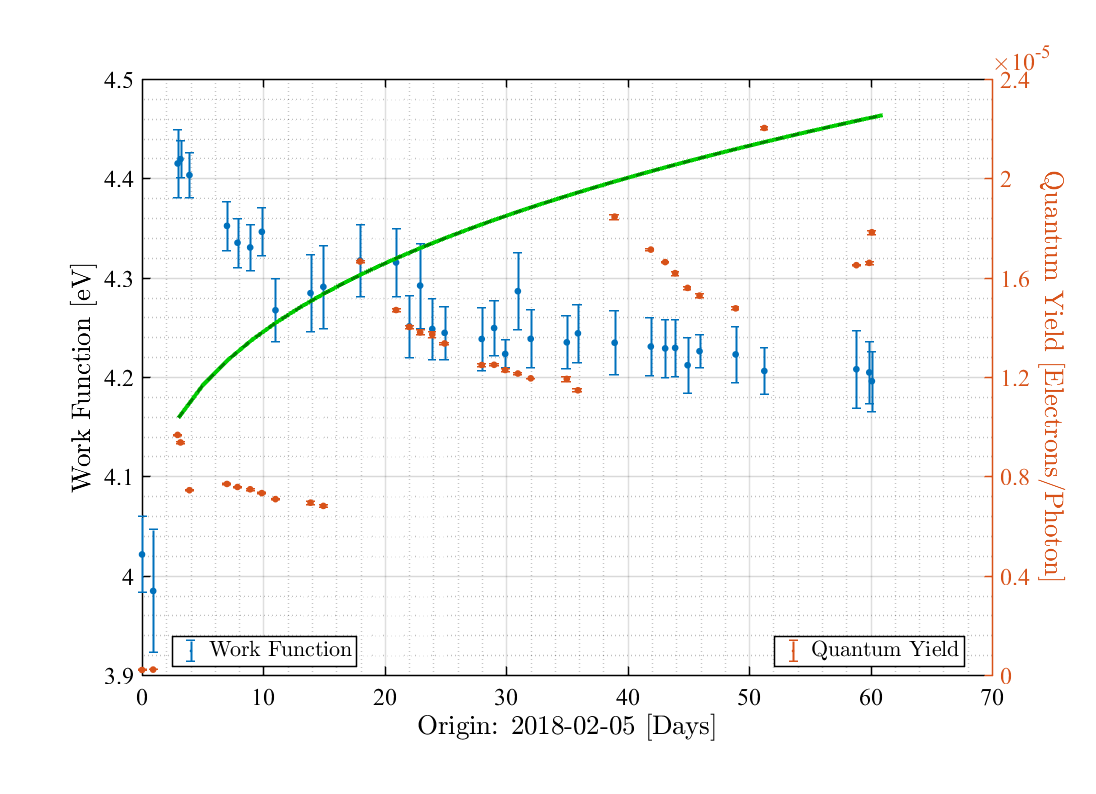}
\caption{\label{MTK800yield2}  Quantum yield from MTK800-01 during a 60 day period in which four bake-out cycles were used to recover $QY_{ext}$ back to its previous equilibrium value.  Also shown are simultaneous work function values (in blue) and an illustrative yield variation (green curve) implied by the work function behaviour}
\end{figure}

Up to this point none of the surfaces had been through any bake-out procedure. So for run 6 a thin film heater was attached to the underside of the unresponsive MTK800-01 and it was reinstalled into the vacuum chamber.   The initial, pre-bake, data still showed no measurable yield until the surface had been in vacuum for nearly 7 days, when a result of $2\times10^{-7}$\,e/photon was then achieved and this is taken as the starting condition. Over the next 60 days the surface was put through four bake-out cycles to $\sim115^o$C.  The results are shown in figure~\ref{MTK800yield2}.  The first bake-out lasted 24\,hrs and the surface was allowed to cool back to room temperature before the a measurement the next day gave a yield close to $10^{-5}$\,e/photon.  The yield then partially relaxed back down by $\sim30\%$ over then 12\,days.  The next three heating cycles each lasted 48\,hours and produced increases in yield followed by partial relaxations.  After the four cycles shown the final equilibrium value is settling down to $\sim$15ppm which is very similar to the equilibrium value for the same surface shown in figure~\ref{yields} (right).   Figure~\ref{MTK800yield2} also shows how the work function evolved during the five heating episodes.   Each heating period lasted a few days and was at a temperature of $\sim115^o$\,C.  After each heating period the quantum yield shows a long time-constant decline which is reminiscent of the behaviour of other, non-heated surfaces.  By contrast with the other work function measurements a trend is seen with a decline of about 0.25eV, which is consistent with the overall trend in the quantum yield increase, as indicated by the green line.  The data are not of sufficient quality to track whether this anti-correlation holds throughout or whether it is purely associated with the heating periods.  The fact that there is a sufficient (anti-)correlation between quantum yield and work function suggests that the first heating has removed the bulk of the `contamination' which was suppressing the emission.  After that, the active area remains constant but there is further reduction in work function as less tightly bound adsorbents are removed.  The relaxation periods between each heating cycle suggests that the heating does indeed disturb the equilibrium distribution of molecular attachments and that these then relax back towards equilibrium with a time constant similar to that shown for the other studies.     

\begin{figure}[h]
\includegraphics[width=3.2in]{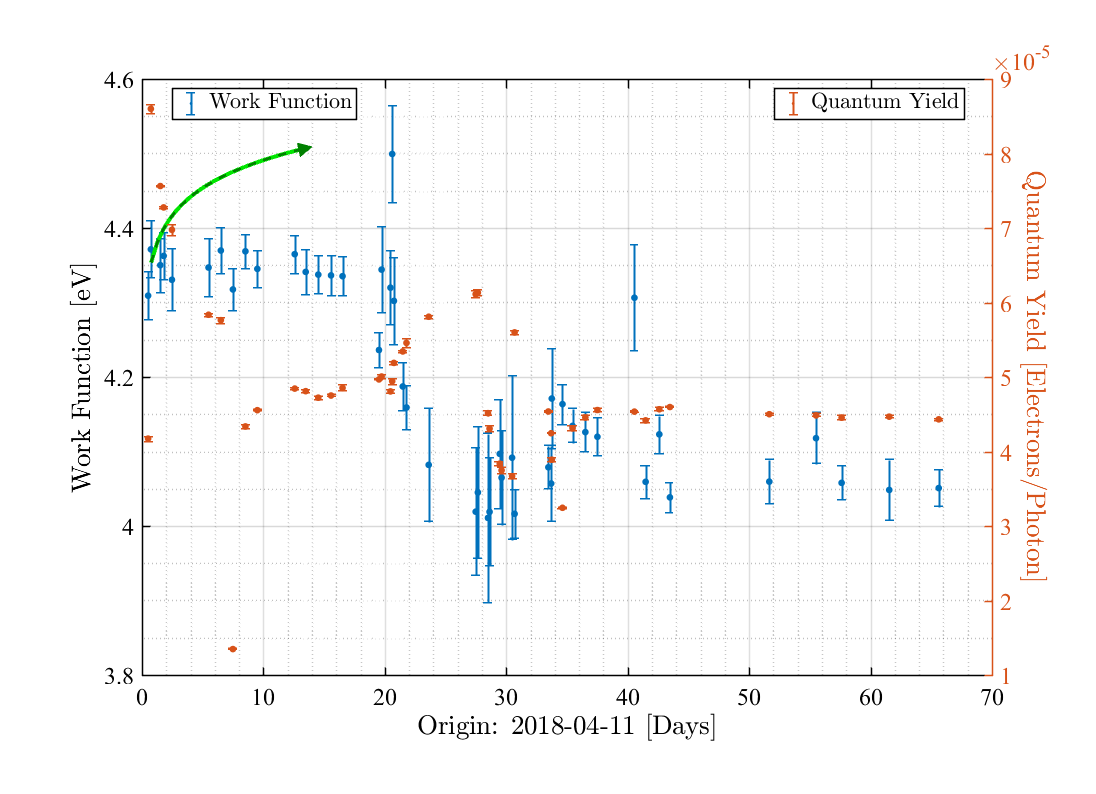}
\includegraphics[width=3.2in]{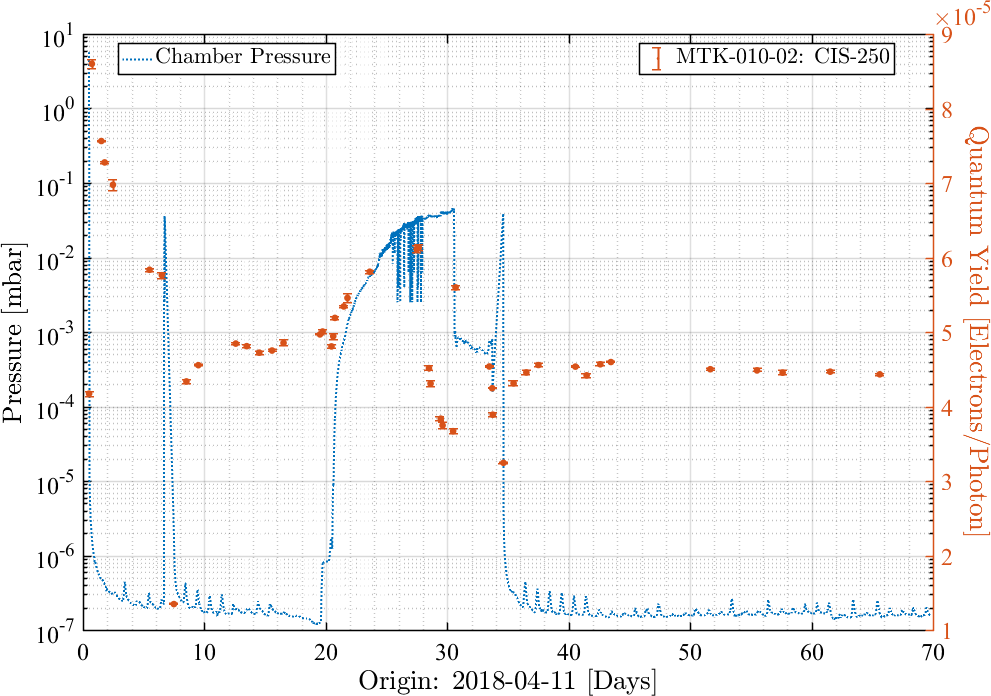}
\caption{\label{MTK0102yield}  (Left) $QY_{ext}$ from MTK010-02 during a 10 week period following first use after 3 years sealed storage.  Work function measurements are in blue. Superimposed (green dashed curve) during the initial decline period, up to $1.5\times10^6$\,s, is an estimate of how the work function would have needed to have changed to explain the quantum yield decline. (Right) The vacuum pressure history is shown. The small quasi-periodic changes are due to daily temperature changes.}
\end{figure}

For run 7 an unused surface, MTK010-02, was recovered from its original sealed (under dry N$_2$) delivery packaging in which it had been for 3 years.  It was placed into vacuum without cleaning and within a few minutes to avoid undue exposure to air and to obtain early measurements of the time evolution.   Figure~\ref{MTK0102yield} shows the results.  Ignoring the initial low measurement taken whilst the chamber pressure was still relatively high, it can be seen that this surface is behaving much like the original surfaces shown in figure~\ref{yields} (right) except that the yield is a factor of a few higher.  In order to investigate the first low measurement point, and one at around $7.5\,$days, which was taken at the time of a power outage, the effect of the vacuum environment has been explored by controlling the chamber pressure.  The results are shown in the figure~\ref{MTK0102yield} (bottom) over a 10 week period.   It can be seen that the very first measurement was taken whilst the chamber was in its initial pump down cycle.  The effect of the power outage can also be seen.  However, the general correlation between the chamber pressure itself and quantum yield is not strong.  After the initial decline the $QY_{ext}$ for this surface is fairly stable at $\sim45$ppm. During the initial decline of the quantum yield, which lasts $\approx10$\,days for this sample, the measured work function is constant with a value $\sim 4.35$\,eV, which is consistent with that measured for the first three samples (and also with that expected for a monolayer of water~\cite{wells72}).  However it is clear from these data that the work function is not changing in a way that would explain the decline.  This is illustrated by the green curve plotted on the left figure. The  necessary increase in work function is $\sim$\,0.15eV.  However, the work function is constant, implying the change in the yield comes from a change in the fraction of active area.    At around $19.5$\,days the pressure in the vacuum chamber was increased, by throttling the vacuum pumping speed.  There is a clear coincident reduction in the work function down to $\sim 4.15$\,eV.  Between $20$ and $28$\,days there is a small, but systematic, increase in the quantum yield by about 20\% which looks to be associated with the work function decrease.  However, if the work function had changed by such a large amount over the whole active emitting surface then equation~\ref{qypatch} predicts an increase closer to a factor of 2.  Hence the implication is that whilst some new (polar) adsorbent has lowered the work function there has also been a non-emissive (non-polar) contaminant which has reduced the effective active area.   Between $28$ and $35$\,days there is some complex behaviour in the quantum yield data as the vacuum conditions became more uncertain (probably due to automatic switching between pressure gauges).  There is a hint of a rapid quantum yield change at $\sim29$\,days followed by a recovering between $30$ and $38$\,days. However after $38$\,days, as the vacuum is re-established, the quantum yield settles down into its equilibrium value, which is only slightly lower than its value around $15$\,days,  but the work function remains at its lower value.   This confirms that the large work function decrease is not due to additional physisorbed water layers, as seen by \cite{wells72}, as this effect would have been reversible.  

\begin{figure}[h]
\includegraphics[width=3.2in]{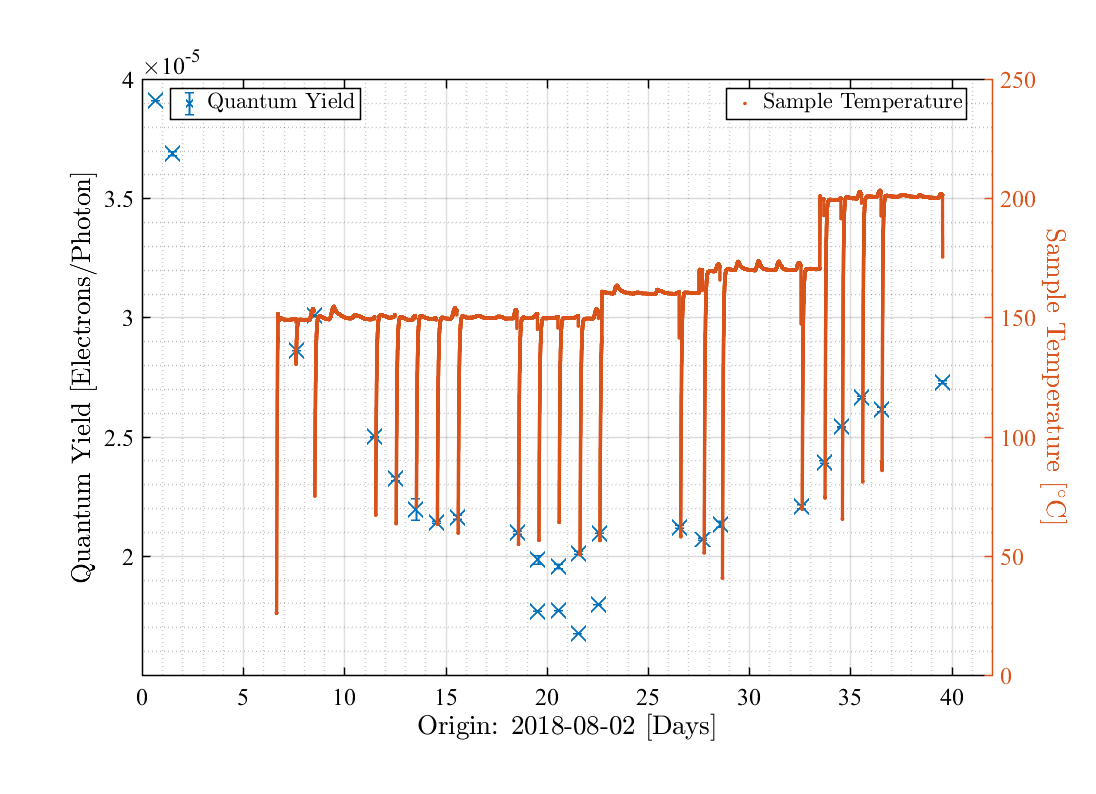}
\includegraphics[width=3.2in]{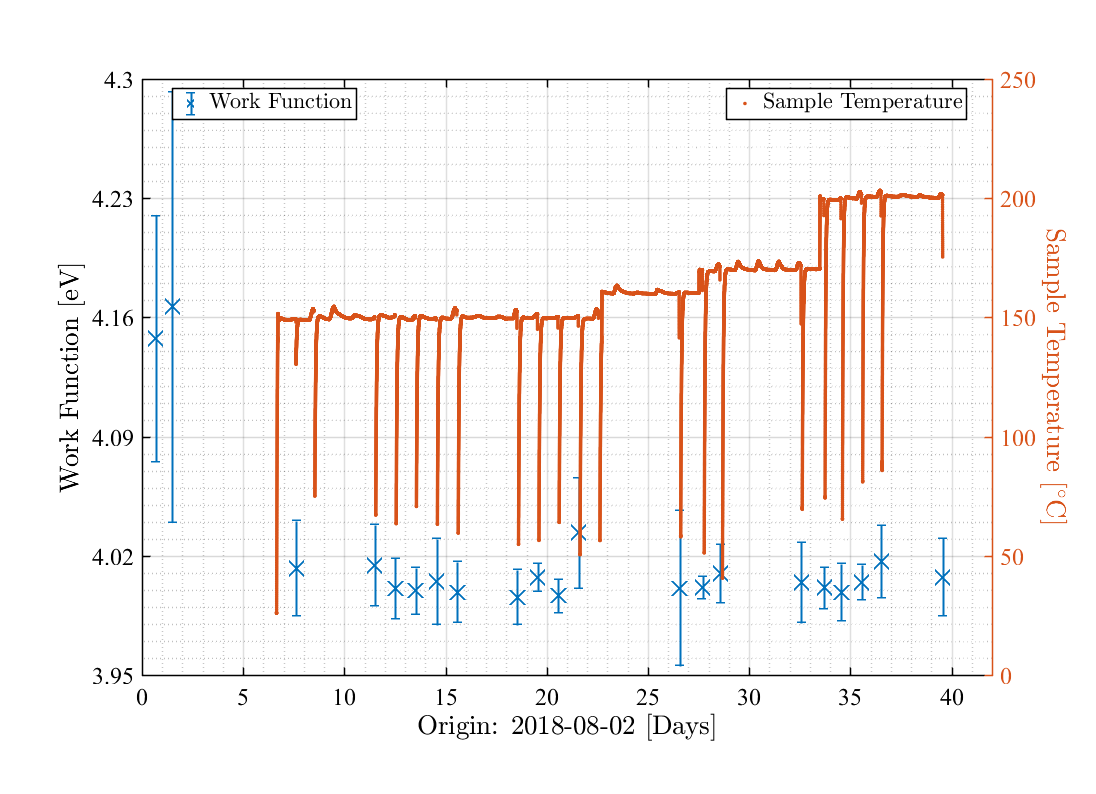}
\caption{\label{MTK010-02_bake}  Results from MTK010-02 during a series of bake-out experiments following on from Figure~\ref{MTK0102yield}. (Left) $QY_{ext}$ and temperature. (Right) Work function and temperature.}
\end{figure}

Following the result from run 7, implying that MTK010-02 was now contaminated, a heater was attached to it to see if any `cleaning' could be effected by bake-out in vacuum.   The initial $QY_{ext}$ measurements, once reinstalled into vacuum for run 8, were within a factor of 2 of those measured before removal from vacuum, and the work function measurements were consistent.  The surface was then subjected to a prolonged heating/bake-out campaign at progressively higher temperatures from $\sim150^\circ$\,C to $\sim200^\circ$\,C.  The yield showed some $\sim30\%$ variation, but not with any overall trend.  The work function rapidly settled down to $\sim 4.0$\,eV and then remained constant.  These results are shown in Figure~\ref{MTK010-02_bake}.  The bake-out did not restore the work function to $\sim 4.3$\,eV, so  it appears that whatever contaminant was present, and was responsible for the photoelectric emission, was non-volatile.

\begin{figure}[h]
\includegraphics[width=3.2in]{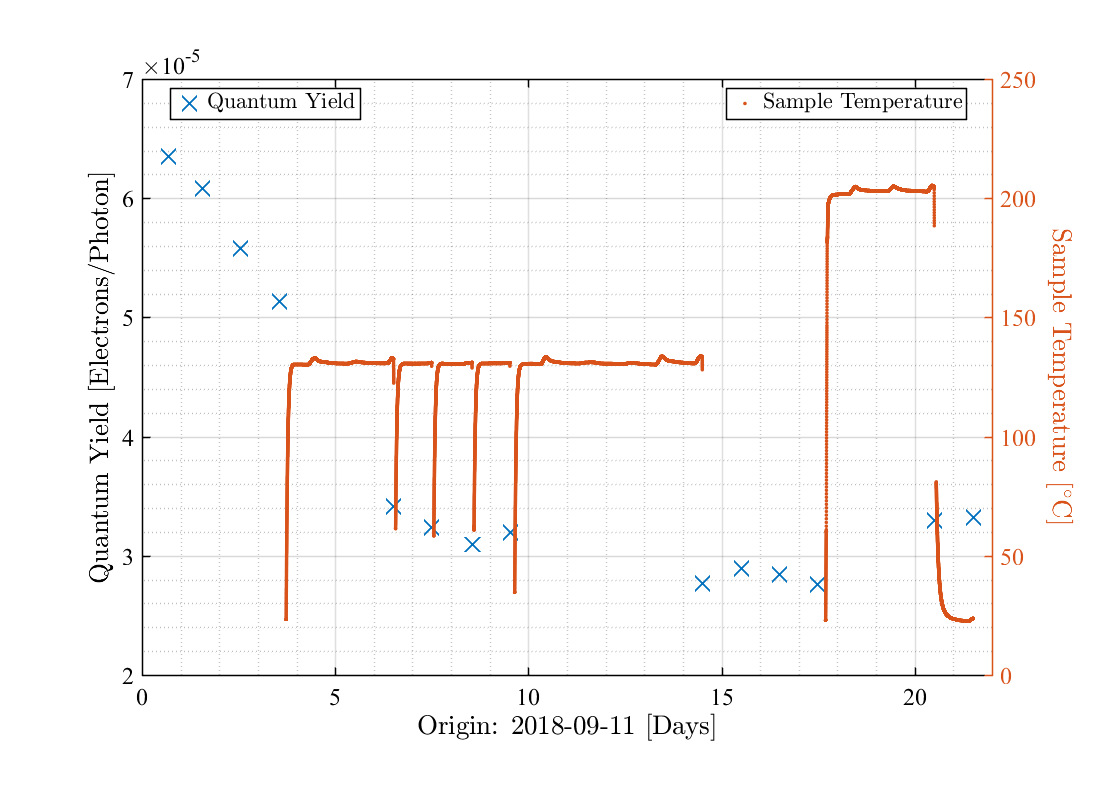}
\includegraphics[width=3.2in]{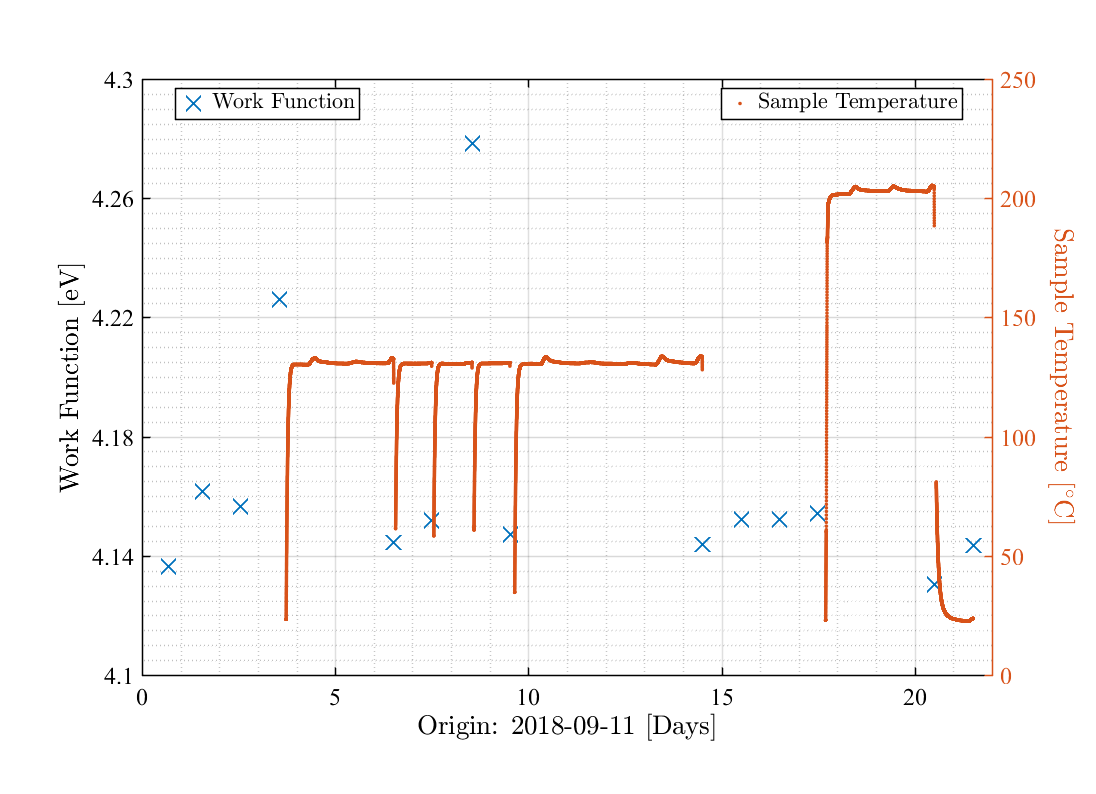}
\caption{\label{MTK010-03_bake}  Quantum yield from MTK010-03 during a series of bake-out experiments in vacuum. The left panel shows the quantum yield and temperature, whilst the right panel shows the work function.}
\end{figure}

For the final run 9,  a fresh, unused, surface, MTK010-03, was installed with an attached heater into the vacuum.   Its initial yield, before any bake-out was high ($\sim 65$ppm) and declined by some 17\% over the first 3 days.  At this point the first bake was done at $130^\circ$\,C.  Over the course of the next 20 days the surface was baked almost continuously at  $130^\circ$\,C and then  $200^\circ$\,C.  The yield showed no obvious response to the high-temperature and continued to decline down to $\sim 45$\% of its original value, whilst the work function remained stable at $\sim 4.16$\,eV, although the results were more scattered than usual, with a single high result close to 4.3.  These results are shown in Figure~\ref{MTK010-03_bake}.  There was no post-bake 'recovery' in the yield.
 
\section{\label{discussion}Discussion and Application}
A comprehensive investigation into the practical behaviour of the photoelectric response of large area gold surfaces under UV illumination has been completed.  The first motivation was to consolidate our understanding of the empirical behaviour to enable accurate similations of how the process will effect the discharging of isolated proof-masses within metallic enclosures in a space environment.  An appropriate model involves a thin (mono-layer) of polar molecules providing a lower single-valued work function surface.   The second motivation was to elucidate the absolute levels of photoelectric emission, including time evolution, from gold-plated surfaces which are manufactured commercially and handled in a controlled clean fashion but without the option of in-situ production/use in vacuum.  This work has shown several, hitherto unknown, phenomenological behaviours from a  modest number of surfaces, but, nonetheless showing the ability to achieve a sufficient final stable yield.

\subsection{The model}
\subsubsection{The basic concept}
A signficant lowering of the work function of gold (from 5.1\,eV to 4.4\,eV and below) occurs by the chemisorption of a tightly bound monolayer of polar molecules, such as water~\cite{wells72},  and/or  hydrocarbons, which exist in both polar and non-polar forms.~\cite{nelson67, hechenblaikner12}  Using such a model, equations~\ref{QYi}, \ref{QYeenergy}, \ref{pesc}, \ref{eangle} and \ref{qypatch} encompass the functional dependencies of photoelectric yield and electron energy distribution on photon energy, angle of incidence, scattering and surface patchiness which are needed for photon ray-tracing, photoelectron production and electron propagation simulations.   A model incorporating this physics has been successfully applied to LISA Pathfinder~\cite{armano18}.
\subsubsection{Time evolution in quantum yield and work function}

Important  aspects from this study are the unexpected long-term time constants in the evolution of the quantum yield, the associated work function behaviour and the efficacy of surface treatments.   Although, it had been anticipated that physisorbed layers, (known to reduce the work function to $\sim$\,3.9\,eV) would be lost through vacuum desorption, the expected timescale is relatively short and incompatible with time dependencies seen in quantum yield, and, more crucially, incompatible with work function behaviours seen in this work.  In all but one case no changes in work function were seen and hence the evolutionary process must be one that changes the active surface area.  The possibilities are: a change in the area coverage of (a) the chemisorbed polar layer(s) or (b) some additional non-emissive surface contaminant, and (c) some change in the effectiveness of the chemisorbed layer.  The mechanism needs to explain approach to similar equilibrium $QY_{ext}$ values from above and below depending on the starting value {\it without any associated change in the work function}, as shown by the surface MTK010-01 in runs 2 and 5 (see figures~\ref{yields}(right) and \ref{MTK010yield2}).  This is hard to explain with (a) or (b) alone.  Although (b) was likely playing a role in the behaviour of MTK800-01 in figure~\ref{MTK800yield2} where the bake-out cycles were needed to restore $QY_{ext}$, it does not explain the relaxation in yield after each cycle. Similarly, whilst (a) might explain figure~\ref{yields} (right), it will not explain figure~\ref{MTK010yield2} where there is a recovery in yield without any bake-out.    (c) is motivated because the attachment of water on a gold surface has many subtleties and can be affected by other species, such as oxygen~\cite{mullen13}, or other contaminants~\cite{bewig65}. There are multiple stable orientations for the first layer water molecules and preferred sites depending on the crystal planes available.~\cite{huzayyin14} The orientation (and presumably site location) can change,~\cite{velasco14}  with binding energies ranging from 0.09 -- 0.19\,eV. The most tightly bound is `hydrogen up' on an Au(110) site, whilst the other fifteen orientation/site combinations are closer to `hydrogen down'.~\cite{huzayyin14}    Wells notes~\cite{wells72} that the `oxygen downward orientation' gives the `positive outward dipole' required to lower the work function. Hence a water molecule migrating from  `hydrogen up' to `hydrogen down' becomes inactive as a mediator for photoelectric emission.  At room temperature there will be some equilibrium distribution of attachment orientations/sites which will leave some fraction of the surface area active in photoelectric emission.   Any process, such as exposure to air or bake-out, which disturbs the distribution  away from equilibrium would be followed by a subsequent relaxation, and this could be in either direction depending on the non-equilibrium starting state.  {\it However}, the binding energies (and their differences) are not compatible with thermal relaxation with the time constant(s) seen in this study.  They are also lower than previously found for the chemisorbed water monolayer of  $\sim25\,$kcal/mol or  $\sim1\,$eV. Indeed, even the higher binding energy for the chemisorbed monolayer is too low to explain the observed time constants, as the associated desorption time is $\sim$ few hours.  To avoid complete loss of the monolayer it must be replenished.  A  water partial pressure of $> \sim10^{-10}$\,mbar would have been sufficient and this is not unreasonable within the vacuum chamber used (or within the LISA Pathfinder gravitational reference sensor~\cite{armano18a}).  However this effect should cause some dependence of both the time constants and the equilibrium quantum yield on the residual partial pressure.  Whilst there is some variation in both within this study, a detailed correlation is not clear.  Note that both partial depletion in the water monolayer coverage and redistribution of the attached water molecule orientation distribution can affect the quantum yield without altering the work function, as observed in Figures~\ref{yields} (left),~\ref{yields} (right) and \ref{MTK0102yield}, and as such is an attractive model.  {\it However}, the time constants remain problematic as the binding energies look too low, and the data for MTK010-02, in Figure~\ref{MTK010-02_bake}, and MTK010-03, in Figure~\ref{MTK010-03_bake}, show no thermal effects up to $200^\circ$\,C.

The data shown in Figures~\ref{MTK010yield2},  \ref{MTK800yield2}, \ref{MTK0102yield}, \ref{MTK010-02_bake} and \ref{MTK010-03_bake}  show significantly different final work functions to that expected for a water monolayer.  A lower work function can be explained by a thin film of contamination, which has either a larger internal polar dipole moment or less internal scattering causing the escaping electrons to loose energy.   The significant  $QY_{ext}$ increase in Figure~\ref{MTK010yield2}, without any decrease in work function could be due to desorption in vacuum at room temperature of a second contaminant species gradually exposing the emissive underlying surface.  Figure~\ref{MTK800yield2} shows a surface which started out in a very heavily contaminated condition with no detectable emission.  With successive heating cycles the surface becomes more and more emissive, and the work function decreases from a high value down to $\sim4.2\,$eV,  in a way which explains the overall increasing trend in the $QY_{ext}$.  However, periods are seen after each heating which show some `relaxation' very similar to that seen in the clean samples.   This could suggest that partial depletion in the water monolayer coverage and/or redistribution of the attached water molecule orientation distribution is being seen, together with progressive removal of a volatile contaminant layer.   In Figure~\ref{MTK0102yield} there is an abrupt change in the work function presumably due to the real-time deposit of a non-volatile thin film contamination, as the vacuum integrity is compromised. It remains in place despite subsequent baking to $\sim200^\circ$\,C.

Table~\ref{summary_behaviours} summarises the behaviours shown by the various surfaces.
\begin{table}[h]
\caption{\label{summary_behaviours}Summary of surface behaviours }
\begin{ruledtabular}
\begin{tabular}{ccccccc}
Surface  & QY$_i$ & QY$_f$ & $\tau$ & WF$_i$ & WF$_f$ & Figs/Tables \\ 
& (ppm) & (ppm) & days & eV & eV & \\ \hline
MTK010-01 & $\sim 40$ & 11 & $\sim 7\downarrow$ & -- & 4.40 & \ref{yields} (right)/\ref{workfs}, \ref{workf}\\
MTK010-01~\footnote{After 27 months in storage} & $1.5$ & 12 & $\sim 17\uparrow$ & 4.05 & 4.05 & \ref{MTK010yield2}\\
MTK010-02 & $90$ & 44~\footnote{Temporary rise to $\sim 65$ during a higher pressure excursion} & $\sim 6\downarrow$ & 4.35 & 4.05~\footnote{Abrupt change during pressure change - otherwise constant} & \ref{MTK0102yield} \\
MTK010-02 & $40$ & 26~\footnote{Includes a $\sim 30\%$ increase seen during the final highest temperature bake at $200^\circ$\,C} & $\sim 6\downarrow$ & 4.0 & 4.0 & \ref{MTK010-02_bake}\\
MTK010-03 & $65$ & 30  & $\sim 6\downarrow$ & 4.15 & 4.15 & \ref{MTK010-03_bake}\\
MTK200-01 & $\sim 45$ & 10 & $\sim 7\downarrow$ & -- & 4.41 & \ref{yields}/ \ref{workfs}, \ref{workf}\\
MTK800-01 & $\sim 40$ & 12 & $\sim 7\downarrow$ & -- & 4.29 & \ref{yields} (right)/ \ref{workfs}, \ref{workf}\\
MTK800-01 & $< 0.1 $ & $\sim 20$~\footnote{Recovery using bake-out periods} & $\sim 17\downarrow$~\footnote{In between bake-out periods} & 4.4 & 4.2 & \ref{MTK800yield2}\\

\end{tabular}
\end{ruledtabular}
\end{table}

\subsection{Surface production}
Surfaces have been produced using commercial supply stages which, from the point of view of an application on LISA, have shown sufficiently reliable and consistent quantum yields and work functions without the need for any extreme post-manufacturing or pre-use cleaning measures.  Clean conditions do need to be maintained and controlled to avoid excessive non-volatile hydrocarbon contamination in particular, although our results suggest some adventitious carbon deposits are inevitable~\cite{barr1995}.  These can be very tightly bound ($\sim 280$\,eV, and hence resilient to bake-out, and can themselves assist the sub-threshold photoelectric emission.  All surfaces investigated have shown stable long-term quantum yields at similar levels and it has been shown that recovery from non-emissive (contaminated) conditions back to nominal levels can be achieved relatively easily.  The LISA Pathfinder experience shows surfaces can remain stable and emissive over year timescales~\cite{armano18}.

Caution is needed in producing reliable surfaces, especially for long term use in space.  

(a) Surface characterisation requires long duration measurements to understand the behaviour.  The origin of long time constants remains obscure.
(b) In some circumstances $QY_{ext}$ stability probably depends on dynamical exchange of the surface water monolayer with ambient water vapour. Required partial pressures of water are very low, but extreme bake-out procedures ($>250^\circ$\,C) in ultra-high vacuum conditions for all relevant surfaces should be avoided.
(c) The role(s) of adventitious carbon remain unclear, but can be both good and bad for photoelectric emission.  Measurement of $\phi$ provides a valuable diagnostic.

\subsection{Light sources}

The behaviour observed has been fully compliant with the simple expectation of equation~\ref{QYeenergy}.   This implies there is no specific reason to favour shorter wavelength light from the point of view of contamination.  Hence the light source selection should be driven by other constraints and requirements derived from the specific mission application~\cite{hollington15, hollington17}.


%
%

%


\begin{acknowledgments}
This work was partly carried out with support from the European Space Agency under contract C4000103768.  Useful discussion are acknowledged with Tobias Ziegler and Patrick Bergner (Airbus), William Weber (Trento University), Taiwo Olatunde and John Conklin (University of Florida).  FY acknowledges support from the China Scholarship Council.
\end{acknowledgments}

\bibliography{seprdc}

\providecommand{\noopsort}[1]{}\providecommand{\singleletter}[1]{#1}%
\begin{thebibliography}{47}%
\makeatletter
\providecommand \@ifxundefined [1]{%
 \@ifx{#1\undefined}
}%
\providecommand \@ifnum [1]{%
 \ifnum #1\expandafter \@firstoftwo
 \else \expandafter \@secondoftwo
 \fi
}%
\providecommand \@ifx [1]{%
 \ifx #1\expandafter \@firstoftwo
 \else \expandafter \@secondoftwo
 \fi
}%
\providecommand \natexlab [1]{#1}%
\providecommand \enquote  [1]{``#1''}%
\providecommand \bibnamefont  [1]{#1}%
\providecommand \bibfnamefont [1]{#1}%
\providecommand \citenamefont [1]{#1}%
\providecommand \href@noop [0]{\@secondoftwo}%
\providecommand \href [0]{\begingroup \@sanitize@url \@href}%
\providecommand \@href[1]{\@@startlink{#1}\@@href}%
\providecommand \@@href[1]{\endgroup#1\@@endlink}%
\providecommand \@sanitize@url [0]{\catcode `\\12\catcode `\$12\catcode
  `\&12\catcode `\#12\catcode `\^12\catcode `\_12\catcode `\%12\relax}%
\providecommand \@@startlink[1]{}%
\providecommand \@@endlink[0]{}%
\providecommand \url  [0]{\begingroup\@sanitize@url \@url }%
\providecommand \@url [1]{\endgroup\@href {#1}{\urlprefix }}%
\providecommand \urlprefix  [0]{URL }%
\providecommand \Eprint [0]{\href }%
\providecommand \doibase [0]{http://dx.doi.org/}%
\providecommand \selectlanguage [0]{\@gobble}%
\providecommand \bibinfo  [0]{\@secondoftwo}%
\providecommand \bibfield  [0]{\@secondoftwo}%
\providecommand \translation [1]{[#1]}%
\providecommand \BibitemOpen [0]{}%
\providecommand \bibitemStop [0]{}%
\providecommand \bibitemNoStop [0]{.\EOS\space}%
\providecommand \EOS [0]{\spacefactor3000\relax}%
\providecommand \BibitemShut  [1]{\csname bibitem#1\endcsname}%
\let\auto@bib@innerbib\@empty
\bibitem [{\citenamefont {Buchman}\ \emph {et~al.}(1995)\citenamefont
  {Buchman}, \citenamefont {Quinn}, \citenamefont {Keiser}, \citenamefont
  {Gill},\ and\ \citenamefont {Sumner}}]{buchman95}%
  \BibitemOpen
  \bibfield  {author} {\bibinfo {author} {\bibfnamefont {S.}~\bibnamefont
  {Buchman}}, \bibinfo {author} {\bibfnamefont {T.}~\bibnamefont {Quinn}},
  \bibinfo {author} {\bibfnamefont {G.~M.}\ \bibnamefont {Keiser}}, \bibinfo
  {author} {\bibfnamefont {D.}~\bibnamefont {Gill}}, \ and\ \bibinfo {author}
  {\bibfnamefont {T.~J.}\ \bibnamefont {Sumner}},\ }\bibfield  {title}
  {\enquote {\bibinfo {title} {{Charge measurement and control for the Gravity
  Probe B gyroscopes}},}\ }\href@noop {} {\bibfield  {journal} {\bibinfo
  {journal} {Rev. Sci. Instrum.}\ }\textbf {\bibinfo {volume} {66}},\ \bibinfo
  {pages} {120--129} (\bibinfo {year} {1995})}\BibitemShut {NoStop}%
\bibitem [{\citenamefont {Saraf}\ \emph {et~al.}(2016)\citenamefont {Saraf}
  \emph {et~al.}}]{saraf16}%
  \BibitemOpen
  \bibfield  {author} {\bibinfo {author} {\bibfnamefont {S.}~\bibnamefont
  {Saraf}} \emph {et~al.},\ }\bibfield  {title} {\enquote {\bibinfo {title}
  {Ground testing and flight demonstration of charge management of insulated
  test masses using {UV-LED} electron photoemission},}\ }\href@noop {}
  {\bibfield  {journal} {\bibinfo  {journal} {Class. Quant. Grav.}\ }\textbf
  {\bibinfo {volume} {33}},\ \bibinfo {pages} {245004} (\bibinfo {year}
  {2016})}\BibitemShut {NoStop}%
\bibitem [{\citenamefont {Armano}\ \emph {et~al.}(2016)\citenamefont {Armano}
  \emph {et~al.}}]{armano16}%
  \BibitemOpen
  \bibfield  {author} {\bibinfo {author} {\bibfnamefont {M.}~\bibnamefont
  {Armano}} \emph {et~al.},\ }\bibfield  {title} {\enquote {\bibinfo {title}
  {Sub femto-g free-fall for space-borne gravitational wave detectors: {LISA
  Pathfinder} results},}\ }\href@noop {} {\bibfield  {journal} {\bibinfo
  {journal} {Phys. Rev. Letts}\ }\textbf {\bibinfo {volume} {116}},\ \bibinfo
  {pages} {231101} (\bibinfo {year} {2016})}\BibitemShut {NoStop}%
\bibitem [{\citenamefont {Sumner}\ \emph {et~al.}(2007)\citenamefont {Sumner}
  \emph {et~al.}}]{sumner07}%
  \BibitemOpen
  \bibfield  {author} {\bibinfo {author} {\bibfnamefont {T.~J.}\ \bibnamefont
  {Sumner}} \emph {et~al.},\ }\bibfield  {title} {\enquote {\bibinfo {title}
  {{STEP (Satellite Test of the Equivalence Principle)}},}\ }\href@noop {}
  {\bibfield  {journal} {\bibinfo  {journal} {Advs. Space. Res.}\ }\textbf
  {\bibinfo {volume} {39}},\ \bibinfo {pages} {254--258} (\bibinfo {year}
  {2007})}\BibitemShut {NoStop}%
\bibitem [{\citenamefont {Sumner}\ \emph {et~al.}(2004)\citenamefont {Sumner},
  \citenamefont {Araujo}, \citenamefont {Davidge}, \citenamefont {Howard},
  \citenamefont {Lee}, \citenamefont {Rochester}, \citenamefont {Shaul},\ and\
  \citenamefont {Wass}}]{sumner04}%
  \BibitemOpen
  \bibfield  {author} {\bibinfo {author} {\bibfnamefont {T.~J.}\ \bibnamefont
  {Sumner}}, \bibinfo {author} {\bibfnamefont {H.}~\bibnamefont {Araujo}},
  \bibinfo {author} {\bibfnamefont {D.}~\bibnamefont {Davidge}}, \bibinfo
  {author} {\bibfnamefont {A.}~\bibnamefont {Howard}}, \bibinfo {author}
  {\bibfnamefont {C.}~\bibnamefont {Lee}}, \bibinfo {author} {\bibfnamefont
  {G.}~\bibnamefont {Rochester}}, \bibinfo {author} {\bibfnamefont
  {D.}~\bibnamefont {Shaul}}, \ and\ \bibinfo {author} {\bibfnamefont
  {P.}~\bibnamefont {Wass}},\ }\bibfield  {title} {\enquote {\bibinfo {title}
  {Description of charging/discharging processes of the {LISA} sensors},}\
  }\href@noop {} {\bibfield  {journal} {\bibinfo  {journal} {Class. Quant.
  Grav.}\ }\textbf {\bibinfo {volume} {21}},\ \bibinfo {pages} {S597--S602}
  (\bibinfo {year} {2004})}\BibitemShut {NoStop}%
\bibitem [{\citenamefont {Sun}\ \emph {et~al.}(2006)\citenamefont {Sun},
  \citenamefont {Allard}, \citenamefont {Buchman}, \citenamefont {Williams},\
  and\ \citenamefont {Byer}}]{sun06}%
  \BibitemOpen
  \bibfield  {author} {\bibinfo {author} {\bibfnamefont {K.-X.}\ \bibnamefont
  {Sun}}, \bibinfo {author} {\bibfnamefont {B.}~\bibnamefont {Allard}},
  \bibinfo {author} {\bibfnamefont {S.}~\bibnamefont {Buchman}}, \bibinfo
  {author} {\bibfnamefont {S.}~\bibnamefont {Williams}}, \ and\ \bibinfo
  {author} {\bibfnamefont {R.~L.}\ \bibnamefont {Byer}},\ }\bibfield  {title}
  {\enquote {\bibinfo {title} {{LED} deep {UV} source for charge management of
  gravitational reference sensors},}\ }\href@noop {} {\bibfield  {journal}
  {\bibinfo  {journal} {Class. Quant. Grav.}\ }\textbf {\bibinfo {volume}
  {23}},\ \bibinfo {pages} {S141--S150} (\bibinfo {year} {2006})}\BibitemShut
  {NoStop}%
\bibitem [{\citenamefont {Pollack}\ \emph {et~al.}(2010)\citenamefont
  {Pollack}, \citenamefont {Turner}, \citenamefont {Schlamminger},
  \citenamefont {Hagedorn},\ and\ \citenamefont {Gundlach}}]{pollack10}%
  \BibitemOpen
  \bibfield  {author} {\bibinfo {author} {\bibfnamefont {S.~E.}\ \bibnamefont
  {Pollack}}, \bibinfo {author} {\bibfnamefont {M.~D.}\ \bibnamefont {Turner}},
  \bibinfo {author} {\bibfnamefont {S.}~\bibnamefont {Schlamminger}}, \bibinfo
  {author} {\bibfnamefont {C.~A.}\ \bibnamefont {Hagedorn}}, \ and\ \bibinfo
  {author} {\bibfnamefont {J.~H.}\ \bibnamefont {Gundlach}},\ }\bibfield
  {title} {\enquote {\bibinfo {title} {Charge management for gravitational-wave
  observatories using {UV LEDs}},}\ }\href@noop {} {\bibfield  {journal}
  {\bibinfo  {journal} {Phys. Rev. D}\ }\textbf {\bibinfo {volume} {81}},\
  \bibinfo {pages} {021101} (\bibinfo {year} {2010})}\BibitemShut {NoStop}%
\bibitem [{\citenamefont {Kawamura}\ \emph {et~al.}(2011)\citenamefont
  {Kawamura} \emph {et~al.}}]{kawamura11}%
  \BibitemOpen
  \bibfield  {author} {\bibinfo {author} {\bibfnamefont {S.}~\bibnamefont
  {Kawamura}} \emph {et~al.},\ }\bibfield  {title} {\enquote {\bibinfo {title}
  {The {Japenese} space gravitational wave antenna: {DECIGO}},}\ }\href@noop {}
  {\bibfield  {journal} {\bibinfo  {journal} {Class. Quant. Grav.}\ }\textbf
  {\bibinfo {volume} {28}},\ \bibinfo {pages} {094011} (\bibinfo {year}
  {2011})}\BibitemShut {NoStop}%
\bibitem [{\citenamefont {Crowder}\ and\ \citenamefont
  {Cornish}(2005)}]{crowder05}%
  \BibitemOpen
  \bibfield  {author} {\bibinfo {author} {\bibfnamefont {J.}~\bibnamefont
  {Crowder}}\ and\ \bibinfo {author} {\bibfnamefont {N.}~\bibnamefont
  {Cornish}},\ }\bibfield  {title} {\enquote {\bibinfo {title} {Beyond {LISA}:
  {Exploring} future gravitational wave missions},}\ }\href@noop {} {\bibfield
  {journal} {\bibinfo  {journal} {Phys. Rev. D}\ }\textbf {\bibinfo {volume}
  {72}},\ \bibinfo {pages} {083005} (\bibinfo {year} {2005})}\BibitemShut
  {NoStop}%
\bibitem [{\citenamefont {Hu}\ \emph {et~al.}(2017)\citenamefont {Hu} \emph
  {et~al.}}]{hu17}%
  \BibitemOpen
  \bibfield  {author} {\bibinfo {author} {\bibfnamefont {W.-R.}\ \bibnamefont
  {Hu}} \emph {et~al.},\ }\bibfield  {title} {\enquote {\bibinfo {title} {The
  {Taiji} program in space for gravitational wave physics and the nature of
  gravity},}\ }\href@noop {} {\bibfield  {journal} {\bibinfo  {journal} {Natl.
  Sci. Rev.}\ }\textbf {\bibinfo {volume} {4}},\ \bibinfo {pages} {685--686}
  (\bibinfo {year} {2017})}\BibitemShut {NoStop}%
\bibitem [{\citenamefont {Luo}\ \emph {et~al.}(2016)\citenamefont {Luo} \emph
  {et~al.}}]{luo16}%
  \BibitemOpen
  \bibfield  {author} {\bibinfo {author} {\bibfnamefont {J.}~\bibnamefont
  {Luo}} \emph {et~al.},\ }\bibfield  {title} {\enquote {\bibinfo {title}
  {{TianQin}: a space-borne gravitational wave detector},}\ }\href@noop {}
  {\bibfield  {journal} {\bibinfo  {journal} {Class. Quant. Grav.}\ }\textbf
  {\bibinfo {volume} {33}},\ \bibinfo {pages} {035010} (\bibinfo {year}
  {2016})}\BibitemShut {NoStop}%
\bibitem [{\citenamefont {Kawamura}\ \emph {et~al.}(2018)\citenamefont
  {Kawamura} \emph {et~al.}}]{kawamura18}%
  \BibitemOpen
  \bibfield  {author} {\bibinfo {author} {\bibfnamefont {S.}~\bibnamefont
  {Kawamura}} \emph {et~al.},\ }\bibfield  {title} {\enquote {\bibinfo {title}
  {Space gravitational-wave antennas: {DECIGO and B-DECIGO}},}\ }\href@noop {}
  {\bibfield  {journal} {\bibinfo  {journal} {Int. J. Mod. Phys. D}\ ,\
  \bibinfo {pages} {1845001}} (\bibinfo {year} {2018})}\BibitemShut {NoStop}%
\bibitem [{\citenamefont {Kaye}\ and\ \citenamefont {Laby}(2017)}]{kaye}%
  \BibitemOpen
  \bibfield  {author} {\bibinfo {author} {\bibfnamefont {G.~W.~C.}\
  \bibnamefont {Kaye}}\ and\ \bibinfo {author} {\bibfnamefont {T.~H.}\
  \bibnamefont {Laby}},\ }\href@noop {} {\emph {\bibinfo {title} {Tables of
  Physical \& Chemical Constants}}}\ (\bibinfo  {publisher} {National Physics
  Laboratory - Online - http://www.kayelaby.npl.co.uk/},\ \bibinfo {year}
  {2017})\BibitemShut {NoStop}%
\bibitem [{\citenamefont {Buchman}\ \emph {et~al.}(2000)\citenamefont
  {Buchman}, \citenamefont {Everitt}, \citenamefont {Parkinson}, \citenamefont
  {Turneaure}, \citenamefont {Brumley}, \citenamefont {Gill}, \citenamefont
  {Keiser},\ and\ \citenamefont {Xiao}}]{buchman00}%
  \BibitemOpen
  \bibfield  {author} {\bibinfo {author} {\bibfnamefont {S.}~\bibnamefont
  {Buchman}}, \bibinfo {author} {\bibfnamefont {C.~W.~F.}\ \bibnamefont
  {Everitt}}, \bibinfo {author} {\bibfnamefont {B.}~\bibnamefont {Parkinson}},
  \bibinfo {author} {\bibfnamefont {J.~P.}\ \bibnamefont {Turneaure}}, \bibinfo
  {author} {\bibfnamefont {R.}~\bibnamefont {Brumley}}, \bibinfo {author}
  {\bibfnamefont {D.}~\bibnamefont {Gill}}, \bibinfo {author} {\bibfnamefont
  {G.~M.}\ \bibnamefont {Keiser}}, \ and\ \bibinfo {author} {\bibfnamefont
  {Y.}~\bibnamefont {Xiao}},\ }\bibfield  {title} {\enquote {\bibinfo {title}
  {{Gyroscopes and charge control for the Relativity Mission Gravity Probe
  B}},}\ }\href@noop {} {\bibfield  {journal} {\bibinfo  {journal} {Advs. Space
  Res.}\ }\textbf {\bibinfo {volume} {25}},\ \bibinfo {pages} {1181--1184}
  (\bibinfo {year} {2000})}\BibitemShut {NoStop}%
\bibitem [{\citenamefont {Ziegler}\ \emph {et~al.}(2014)\citenamefont
  {Ziegler}, \citenamefont {Bergner}, \citenamefont {Hechenblaikner},
  \citenamefont {Brandt},\ and\ \citenamefont {Fichter}}]{ziegler14}%
  \BibitemOpen
  \bibfield  {author} {\bibinfo {author} {\bibfnamefont {T.}~\bibnamefont
  {Ziegler}}, \bibinfo {author} {\bibfnamefont {P.}~\bibnamefont {Bergner}},
  \bibinfo {author} {\bibfnamefont {G.}~\bibnamefont {Hechenblaikner}},
  \bibinfo {author} {\bibfnamefont {N.}~\bibnamefont {Brandt}}, \ and\ \bibinfo
  {author} {\bibfnamefont {W.}~\bibnamefont {Fichter}},\ }\bibfield  {title}
  {\enquote {\bibinfo {title} {Modeling and performance of contact-free
  discharge systems for space inertial sensors},}\ }\href@noop {} {\bibfield
  {journal} {\bibinfo  {journal} {IEEE Trans. Aeros. Elec. Sys.}\ }\textbf
  {\bibinfo {volume} {50}},\ \bibinfo {pages} {1493--1510} (\bibinfo {year}
  {2014})}\BibitemShut {NoStop}%
\bibitem [{\citenamefont {Buchman}\ \emph {et~al.}(2015)\citenamefont
  {Buchman}, \citenamefont {Lipa}, \citenamefont {Keiser}, \citenamefont
  {Muhlfelder},\ and\ \citenamefont {Turneaure}}]{buchman15}%
  \BibitemOpen
  \bibfield  {author} {\bibinfo {author} {\bibfnamefont {S.}~\bibnamefont
  {Buchman}}, \bibinfo {author} {\bibfnamefont {J.~A.}\ \bibnamefont {Lipa}},
  \bibinfo {author} {\bibfnamefont {G.~M.}\ \bibnamefont {Keiser}}, \bibinfo
  {author} {\bibfnamefont {B.}~\bibnamefont {Muhlfelder}}, \ and\ \bibinfo
  {author} {\bibfnamefont {J.~P.}\ \bibnamefont {Turneaure}},\ }\bibfield
  {title} {\enquote {\bibinfo {title} {{The Gravity Probe B gyroscope}},}\
  }\href@noop {} {\bibfield  {journal} {\bibinfo  {journal} {Class. Quant.
  Grav.}\ }\textbf {\bibinfo {volume} {32}},\ \bibinfo {pages} {224004}
  (\bibinfo {year} {2015})}\BibitemShut {NoStop}%
\bibitem [{\citenamefont {Armano}\ \emph
  {et~al.}(2018{\natexlab{a}})\citenamefont {Armano} \emph
  {et~al.}}]{armano18}%
  \BibitemOpen
  \bibfield  {author} {\bibinfo {author} {\bibfnamefont {M.}~\bibnamefont
  {Armano}} \emph {et~al.},\ }\bibfield  {title} {\enquote {\bibinfo {title}
  {Precision charge control system for isolated free-falling test masses: {LISA
  Pathfinder} results},}\ }\href@noop {} {\bibfield  {journal} {\bibinfo
  {journal} {Phys. Rev. D}\ }\textbf {\bibinfo {volume} {98}},\ \bibinfo
  {pages} {062001} (\bibinfo {year} {2018}{\natexlab{a}})}\BibitemShut
  {NoStop}%
\bibitem [{\citenamefont {Amaro-Seoane}\ \emph {et~al.}(2017)\citenamefont
  {Amaro-Seoane} \emph {et~al.}}]{Amaro17}%
  \BibitemOpen
  \bibfield  {author} {\bibinfo {author} {\bibfnamefont {P.}~\bibnamefont
  {Amaro-Seoane}} \emph {et~al.},\ }\bibfield  {title} {\enquote {\bibinfo
  {title} {{Laser Interferometer Space Antenna}},}\ }\href@noop {} {\bibfield
  {journal} {\bibinfo  {journal} {arXiv:1702.00786}\ } (\bibinfo {year}
  {2017})}\BibitemShut {NoStop}%
\bibitem [{\citenamefont {Wells}\ and\ \citenamefont {{T. Fort
  Jr.}}(1972)}]{wells72}%
  \BibitemOpen
  \bibfield  {author} {\bibinfo {author} {\bibfnamefont {R.~L.}\ \bibnamefont
  {Wells}}\ and\ \bibinfo {author} {\bibnamefont {{T. Fort Jr.}}},\ }\bibfield
  {title} {\enquote {\bibinfo {title} {Adsoption of water on clean gold by
  measurement of work function changes},}\ }\href@noop {} {\bibfield  {journal}
  {\bibinfo  {journal} {Sur. Sci.}\ }\textbf {\bibinfo {volume} {32}},\
  \bibinfo {pages} {554--560} (\bibinfo {year} {1972})}\BibitemShut {NoStop}%
\bibitem [{\citenamefont {Barr}\ and\ \citenamefont {Seal}(1995)}]{barr1995}%
  \BibitemOpen
  \bibfield  {author} {\bibinfo {author} {\bibfnamefont {T.~L.}\ \bibnamefont
  {Barr}}\ and\ \bibinfo {author} {\bibfnamefont {S.}~\bibnamefont {Seal}},\
  }\bibfield  {title} {\enquote {\bibinfo {title} {Nature of the use of
  adventitious carbon as a binding energy standard},}\ }\href@noop {}
  {\bibfield  {journal} {\bibinfo  {journal} {J. Vac. Sci. \& Techn. A}\
  }\textbf {\bibinfo {volume} {13}},\ \bibinfo {pages} {1239--1246} (\bibinfo
  {year} {1995})}\BibitemShut {NoStop}%
\bibitem [{\citenamefont {Alloway}\ \emph {et~al.}(2009)\citenamefont
  {Alloway}, \citenamefont {Graham}, \citenamefont {Yang}, \citenamefont
  {Mudalige}, \citenamefont {{Colorado Jr.}}, \citenamefont {Wysocki},
  \citenamefont {Pemberton}, \citenamefont {Lee}, \citenamefont {Wysocki},\
  and\ \citenamefont {Armstrong}}]{alloway09}%
  \BibitemOpen
  \bibfield  {author} {\bibinfo {author} {\bibfnamefont {D.~M.}\ \bibnamefont
  {Alloway}}, \bibinfo {author} {\bibfnamefont {A.~L.}\ \bibnamefont {Graham}},
  \bibinfo {author} {\bibfnamefont {X.}~\bibnamefont {Yang}}, \bibinfo {author}
  {\bibfnamefont {A.}~\bibnamefont {Mudalige}}, \bibinfo {author}
  {\bibfnamefont {R.}~\bibnamefont {{Colorado Jr.}}}, \bibinfo {author}
  {\bibfnamefont {V.~H.}\ \bibnamefont {Wysocki}}, \bibinfo {author}
  {\bibfnamefont {J.~E.}\ \bibnamefont {Pemberton}}, \bibinfo {author}
  {\bibfnamefont {T.~R.}\ \bibnamefont {Lee}}, \bibinfo {author} {\bibfnamefont
  {R.~J.}\ \bibnamefont {Wysocki}}, \ and\ \bibinfo {author} {\bibfnamefont
  {N.~R.}\ \bibnamefont {Armstrong}},\ }\bibfield  {title} {\enquote {\bibinfo
  {title} {Tuning the effective work function of gold and silver using
  w-functionalized alkanethaniols: Varying surface composition through dilution
  and choice of terminal groups},}\ }\href@noop {} {\bibfield  {journal}
  {\bibinfo  {journal} {J. Phys. Chem.}\ }\textbf {\bibinfo {volume} {113}},\
  \bibinfo {pages} {20328--20334} (\bibinfo {year} {2009})}\BibitemShut
  {NoStop}%
\bibitem [{\citenamefont {O'Hanlon}(2005)}]{hanlon05}%
  \BibitemOpen
  \bibfield  {author} {\bibinfo {author} {\bibfnamefont {J.~F.}\ \bibnamefont
  {O'Hanlon}},\ }\href@noop {} {\emph {\bibinfo {title} {{A User's Guide to
  Vacuum Technology, $3^{\rm rd}$ Edition}}}}\ (\bibinfo  {publisher} {Wiley},\
  \bibinfo {year} {2005})\BibitemShut {NoStop}%
\bibitem [{\citenamefont {Saville}\ \emph {et~al.}(1995)\citenamefont
  {Saville}, \citenamefont {Platzman}, \citenamefont {Brandes}, \citenamefont
  {Ruel},\ and\ \citenamefont {Willett}}]{saville95}%
  \BibitemOpen
  \bibfield  {author} {\bibinfo {author} {\bibfnamefont {G.~F.}\ \bibnamefont
  {Saville}}, \bibinfo {author} {\bibfnamefont {P.~M.}\ \bibnamefont
  {Platzman}}, \bibinfo {author} {\bibfnamefont {G.}~\bibnamefont {Brandes}},
  \bibinfo {author} {\bibfnamefont {R.}~\bibnamefont {Ruel}}, \ and\ \bibinfo
  {author} {\bibfnamefont {R.~L.}\ \bibnamefont {Willett}},\ }\href@noop {}
  {\enquote {\bibinfo {title} {Feasibility study of photocathode electron
  projection lithography},}\ } (\bibinfo {year} {1995})\BibitemShut {NoStop}%
\bibitem [{\citenamefont {Jiang}\ \emph {et~al.}(1998)\citenamefont {Jiang},
  \citenamefont {Berglund}, \citenamefont {Bell},\ and\ \citenamefont
  {Mackie}}]{jiang98}%
  \BibitemOpen
  \bibfield  {author} {\bibinfo {author} {\bibfnamefont {X.}~\bibnamefont
  {Jiang}}, \bibinfo {author} {\bibfnamefont {C.~N.}\ \bibnamefont {Berglund}},
  \bibinfo {author} {\bibfnamefont {A.~E.}\ \bibnamefont {Bell}}, \ and\
  \bibinfo {author} {\bibfnamefont {W.~A.}\ \bibnamefont {Mackie}},\ }\bibfield
   {title} {\enquote {\bibinfo {title} {Photoemission from gold thin films for
  application in multiphotocathode arrays for electron beam lithography},}\
  }\href@noop {} {\bibfield  {journal} {\bibinfo  {journal} {J. Vac. Sci.
  Technol. B}\ }\textbf {\bibinfo {volume} {16}},\ \bibinfo {pages}
  {3374--3379} (\bibinfo {year} {1998})}\BibitemShut {NoStop}%
\bibitem [{\citenamefont {Smith}\ \emph {et~al.}(1974)\citenamefont {Smith},
  \citenamefont {Wertheim}, \citenamefont {Hufner},\ and\ \citenamefont
  {Traum}}]{smith74}%
  \BibitemOpen
  \bibfield  {author} {\bibinfo {author} {\bibfnamefont {N.~V.}\ \bibnamefont
  {Smith}}, \bibinfo {author} {\bibfnamefont {G.~K.}\ \bibnamefont {Wertheim}},
  \bibinfo {author} {\bibfnamefont {S.}~\bibnamefont {Hufner}}, \ and\ \bibinfo
  {author} {\bibfnamefont {M.~M.}\ \bibnamefont {Traum}},\ }\bibfield  {title}
  {\enquote {\bibinfo {title} {{Photoemission spectra and band structures of
  d-band metals. IV. X-ray photoemission spectra and densities of states in Rh,
  Pd, Ag, Ir, Pt, and Au}},}\ }\href@noop {} {\bibfield  {journal} {\bibinfo
  {journal} {Phys. Rev. B}\ }\textbf {\bibinfo {volume} {10}},\ \bibinfo
  {pages} {3197} (\bibinfo {year} {1974})}\BibitemShut {NoStop}%
\bibitem [{\citenamefont {Johnson}\ and\ \citenamefont
  {Christy}(1972)}]{johnson72}%
  \BibitemOpen
  \bibfield  {author} {\bibinfo {author} {\bibfnamefont {P.~B.}\ \bibnamefont
  {Johnson}}\ and\ \bibinfo {author} {\bibfnamefont {R.~W.}\ \bibnamefont
  {Christy}},\ }\bibfield  {title} {\enquote {\bibinfo {title} {Optical
  constants of the noble metals},}\ }\href@noop {} {\bibfield  {journal}
  {\bibinfo  {journal} {Phys. Rev. B}\ }\textbf {\bibinfo {volume} {6}},\
  \bibinfo {pages} {4370--4379} (\bibinfo {year} {1972})}\BibitemShut {NoStop}%
\bibitem [{\citenamefont {Zombeck}(1990)}]{zombeck90}%
  \BibitemOpen
  \bibfield  {author} {\bibinfo {author} {\bibfnamefont {M.~V.}\ \bibnamefont
  {Zombeck}},\ }\href@noop {} {\emph {\bibinfo {title} {Handbook of Space
  Astronomy \& Astrophysics}}}\ (\bibinfo  {publisher} {Cambridge University
  Press},\ \bibinfo {year} {1990})\BibitemShut {NoStop}%
\bibitem [{\citenamefont {Hollington}(2011)}]{hollington11}%
  \BibitemOpen
  \bibfield  {author} {\bibinfo {author} {\bibfnamefont {D.}~\bibnamefont
  {Hollington}},\ }\emph {\bibinfo {title} {The charge management system for
  {LISA and LISA Pathfinder}}},\ \href@noop {} {\bibinfo {type} {{Ph.D.}
  thesis}},\ \bibinfo  {school} {Imperial College London} (\bibinfo {year}
  {2011})\BibitemShut {NoStop}%
\bibitem [{\citenamefont {Adams}\ \emph {et~al.}(1987)\citenamefont {Adams},
  \citenamefont {Rochester}, \citenamefont {Sumner},\ and\ \citenamefont
  {Williams}}]{adams87}%
  \BibitemOpen
  \bibfield  {author} {\bibinfo {author} {\bibfnamefont {G.~P.}\ \bibnamefont
  {Adams}}, \bibinfo {author} {\bibfnamefont {G.~K.}\ \bibnamefont
  {Rochester}}, \bibinfo {author} {\bibfnamefont {T.~J.}\ \bibnamefont
  {Sumner}}, \ and\ \bibinfo {author} {\bibfnamefont {O.~R.}\ \bibnamefont
  {Williams}},\ }\bibfield  {title} {\enquote {\bibinfo {title} {The
  ultraviolet calibration system for the {UK XUV} telescope to be flown on the
  {ROSAT} satellite},}\ }\href@noop {} {\bibfield  {journal} {\bibinfo
  {journal} {J. Phys. E: Sci. Instrum.}\ }\textbf {\bibinfo {volume} {20}},\
  \bibinfo {pages} {1261--1264} (\bibinfo {year} {1987})}\BibitemShut {NoStop}%
\bibitem [{\citenamefont {Wass}\ \emph {et~al.}(2006)\citenamefont {Wass} \emph
  {et~al.}}]{wass06}%
  \BibitemOpen
  \bibfield  {author} {\bibinfo {author} {\bibfnamefont {P.~J.}\ \bibnamefont
  {Wass}} \emph {et~al.},\ }\bibfield  {title} {\enquote {\bibinfo {title}
  {Testing of the {UV} discharge system for {LISA Pathfinder}},}\ }\href@noop
  {} {\bibfield  {journal} {\bibinfo  {journal} {AIP Conf. Proc.}\ }\textbf
  {\bibinfo {volume} {873}},\ \bibinfo {pages} {220--224} (\bibinfo {year}
  {2006})}\BibitemShut {NoStop}%
\bibitem [{\citenamefont {Olatunde}\ \emph {et~al.}(2015)\citenamefont
  {Olatunde}, \citenamefont {Shelley}, \citenamefont {Chilton}, \citenamefont
  {Serra}, \citenamefont {Ciani}, \citenamefont {Mueller},\ and\ \citenamefont
  {Conklin}}]{taiwo15}%
  \BibitemOpen
  \bibfield  {author} {\bibinfo {author} {\bibfnamefont {T.}~\bibnamefont
  {Olatunde}}, \bibinfo {author} {\bibfnamefont {R.}~\bibnamefont {Shelley}},
  \bibinfo {author} {\bibfnamefont {A.}~\bibnamefont {Chilton}}, \bibinfo
  {author} {\bibfnamefont {P.}~\bibnamefont {Serra}}, \bibinfo {author}
  {\bibfnamefont {G.}~\bibnamefont {Ciani}}, \bibinfo {author} {\bibfnamefont
  {G.}~\bibnamefont {Mueller}}, \ and\ \bibinfo {author} {\bibfnamefont
  {J.}~\bibnamefont {Conklin}},\ }\bibfield  {title} {\enquote {\bibinfo
  {title} {240nm {UV LEDs for LISA} test mass charge control},}\ }\href@noop {}
  {\bibfield  {journal} {\bibinfo  {journal} {J. Phys. Conf. Ser.}\ }\textbf
  {\bibinfo {volume} {610}},\ \bibinfo {pages} {012034} (\bibinfo {year}
  {2015})}\BibitemShut {NoStop}%
\bibitem [{\citenamefont {Hollington}\ \emph {et~al.}(2015)\citenamefont
  {Hollington}, \citenamefont {Baird}, \citenamefont {Sumner},\ and\
  \citenamefont {Wass}}]{hollington15}%
  \BibitemOpen
  \bibfield  {author} {\bibinfo {author} {\bibfnamefont {D.}~\bibnamefont
  {Hollington}}, \bibinfo {author} {\bibfnamefont {J.~T.}\ \bibnamefont
  {Baird}}, \bibinfo {author} {\bibfnamefont {T.~J.}\ \bibnamefont {Sumner}}, \
  and\ \bibinfo {author} {\bibfnamefont {P.~J.}\ \bibnamefont {Wass}},\
  }\bibfield  {title} {\enquote {\bibinfo {title} {Characterising and testing
  deep {UV LEDs} for use in space applications},}\ }\href@noop {} {\bibfield
  {journal} {\bibinfo  {journal} {Class. Quant. Grav.}\ }\textbf {\bibinfo
  {volume} {32}},\ \bibinfo {pages} {235020} (\bibinfo {year}
  {2015})}\BibitemShut {NoStop}%
\bibitem [{\citenamefont {Hollington}\ \emph {et~al.}(2017)\citenamefont
  {Hollington}, \citenamefont {Baird}, \citenamefont {Sumner},\ and\
  \citenamefont {Wass}}]{hollington17}%
  \BibitemOpen
  \bibfield  {author} {\bibinfo {author} {\bibfnamefont {D.}~\bibnamefont
  {Hollington}}, \bibinfo {author} {\bibfnamefont {J.~T.}\ \bibnamefont
  {Baird}}, \bibinfo {author} {\bibfnamefont {T.~J.}\ \bibnamefont {Sumner}}, \
  and\ \bibinfo {author} {\bibfnamefont {P.~J.}\ \bibnamefont {Wass}},\
  }\bibfield  {title} {\enquote {\bibinfo {title} {Lifetime testing {UV LEDs}
  for use in the {LISA} charge management system},}\ }\href@noop {} {\bibfield
  {journal} {\bibinfo  {journal} {Class. Quant. Grav.}\ }\textbf {\bibinfo
  {volume} {34}},\ \bibinfo {pages} {205009} (\bibinfo {year}
  {2017})}\BibitemShut {NoStop}%
\bibitem [{Note1()}]{Note1}%
  \BibitemOpen
  \bibinfo {note} {Http://refractiveindex.info}\BibitemShut {NoStop}%
\bibitem [{\citenamefont {Somorjai}(1981)}]{Somorjai81}%
  \BibitemOpen
  \bibfield  {author} {\bibinfo {author} {\bibfnamefont {G.~A.}\ \bibnamefont
  {Somorjai}},\ }\href@noop {} {\emph {\bibinfo {title} {Chemistry in Two
  Dimensions}}}\ (\bibinfo  {publisher} {Cornell University Press},\ \bibinfo
  {year} {1981})\BibitemShut {NoStop}%
\bibitem [{\citenamefont {Hechenblaikner}\ \emph {et~al.}(2012)\citenamefont
  {Hechenblaikner}, \citenamefont {Ziegler}, \citenamefont {Biswas},
  \citenamefont {Seibel}, \citenamefont {Schulze}, \citenamefont {Brandt},
  \citenamefont {Scholl}, \citenamefont {Berner},\ and\ \citenamefont
  {Reinert}}]{hechenblaikner12}%
  \BibitemOpen
  \bibfield  {author} {\bibinfo {author} {\bibfnamefont {G.}~\bibnamefont
  {Hechenblaikner}}, \bibinfo {author} {\bibfnamefont {T.}~\bibnamefont
  {Ziegler}}, \bibinfo {author} {\bibfnamefont {I.}~\bibnamefont {Biswas}},
  \bibinfo {author} {\bibfnamefont {C.}~\bibnamefont {Seibel}}, \bibinfo
  {author} {\bibfnamefont {M.}~\bibnamefont {Schulze}}, \bibinfo {author}
  {\bibfnamefont {N.}~\bibnamefont {Brandt}}, \bibinfo {author} {\bibfnamefont
  {A.}~\bibnamefont {Scholl}}, \bibinfo {author} {\bibfnamefont
  {P.}~\bibnamefont {Berner}}, \ and\ \bibinfo {author} {\bibfnamefont {F.~T.}\
  \bibnamefont {Reinert}},\ }\bibfield  {title} {\enquote {\bibinfo {title}
  {Energy distribution and quantum yield for photoemission from
  air-contaminated gold surfaces under ultraviolet illumination close to the
  threshold},}\ }\href@noop {} {\bibfield  {journal} {\bibinfo  {journal} {J.
  Appl. Phys.}\ }\textbf {\bibinfo {volume} {111}},\ \bibinfo {pages} {124914}
  (\bibinfo {year} {2012})}\BibitemShut {NoStop}%
\bibitem [{\citenamefont {Pei}\ and\ \citenamefont {Berglund}(2002)}]{pei02}%
  \BibitemOpen
  \bibfield  {author} {\bibinfo {author} {\bibfnamefont {Z.}~\bibnamefont
  {Pei}}\ and\ \bibinfo {author} {\bibfnamefont {C.~N.}\ \bibnamefont
  {Berglund}},\ }\bibfield  {title} {\enquote {\bibinfo {title} {Angular
  distribution of photoemission from gold thin films},}\ }\href@noop {}
  {\bibfield  {journal} {\bibinfo  {journal} {Jap. J. Appl. Phys.}\ }\textbf
  {\bibinfo {volume} {41}},\ \bibinfo {pages} {L52--L54} (\bibinfo {year}
  {2002})}\BibitemShut {NoStop}%
\bibitem [{Note2()}]{Note2}%
  \BibitemOpen
  \bibinfo {note} {Mateck: http://www.mateck.com/}\BibitemShut {NoStop}%
\bibitem [{Note3()}]{Note3}%
  \BibitemOpen
  \bibinfo {note} {Teer Coatings: http://www.teercoatings.co.uk/}\BibitemShut
  {NoStop}%
\bibitem [{Note4()}]{Note4}%
  \BibitemOpen
  \bibinfo {note} {Https://www.alconox.com}\BibitemShut {NoStop}%
\bibitem [{\citenamefont {Ziegler}\ \emph {et~al.}(2013)\citenamefont
  {Ziegler}, \citenamefont {Bergner}, \citenamefont {Hechenblaikner},\ and\
  \citenamefont {Brandt}}]{ziegler13}%
  \BibitemOpen
  \bibfield  {author} {\bibinfo {author} {\bibfnamefont {T.}~\bibnamefont
  {Ziegler}}, \bibinfo {author} {\bibfnamefont {P.}~\bibnamefont {Bergner}},
  \bibinfo {author} {\bibfnamefont {G.}~\bibnamefont {Hechenblaikner}}, \ and\
  \bibinfo {author} {\bibfnamefont {N.}~\bibnamefont {Brandt}},\ }\bibfield
  {title} {\enquote {\bibinfo {title} {{LISA Pathfinder Discharge Working
  Group: Activities, Results, and Lessons Learned for LISA/NGO}},}\ }\href@noop
  {} {\bibfield  {journal} {\bibinfo  {journal} {ASP Conf. Ser.}\ }\textbf
  {\bibinfo {volume} {467}},\ \bibinfo {pages} {317} (\bibinfo {year}
  {2013})}\BibitemShut {NoStop}%
\bibitem [{\citenamefont {{Nelson Jr.}}, \citenamefont {{Lide Jr.}},\ and\
  \citenamefont {Maryott}(1967)}]{nelson67}%
  \BibitemOpen
  \bibfield  {author} {\bibinfo {author} {\bibfnamefont {R.~D.}\ \bibnamefont
  {{Nelson Jr.}}}, \bibinfo {author} {\bibfnamefont {D.~R.}\ \bibnamefont
  {{Lide Jr.}}}, \ and\ \bibinfo {author} {\bibfnamefont {A.~A.}\ \bibnamefont
  {Maryott}},\ }\bibfield  {title} {\enquote {\bibinfo {title} {Selected values
  of electric dipole moments for molecules in the gas phase},}\ }\href@noop {}
  {\bibfield  {journal} {\bibinfo  {journal} {NSRDS-NBS}\ }\textbf {\bibinfo
  {volume} {10}},\ \bibinfo {pages} {1--49} (\bibinfo {year}
  {1967})}\BibitemShut {NoStop}%
\bibitem [{\citenamefont {Mullen}\ \emph {et~al.}(2013)\citenamefont {Mullen},
  \citenamefont {Gong}, \citenamefont {Yang}, \citenamefont {Pan},\ and\
  \citenamefont {{C. Buddie Mullins}}}]{mullen13}%
  \BibitemOpen
  \bibfield  {author} {\bibinfo {author} {\bibfnamefont {G.~M.}\ \bibnamefont
  {Mullen}}, \bibinfo {author} {\bibfnamefont {J.}~\bibnamefont {Gong}},
  \bibinfo {author} {\bibfnamefont {T.}~\bibnamefont {Yang}}, \bibinfo {author}
  {\bibfnamefont {M.}~\bibnamefont {Pan}}, \ and\ \bibinfo {author}
  {\bibnamefont {{C. Buddie Mullins}}},\ }\bibfield  {title} {\enquote
  {\bibinfo {title} {The effects of absorbed water on gold catalysis and
  surface chemistry},}\ }\href@noop {} {\bibfield  {journal} {\bibinfo
  {journal} {Topics in Catalysis}\ }\textbf {\bibinfo {volume} {56}},\ \bibinfo
  {pages} {1499--1511} (\bibinfo {year} {2013})}\BibitemShut {NoStop}%
\bibitem [{\citenamefont {Bewig}\ and\ \citenamefont {Zisman}(1965)}]{bewig65}%
  \BibitemOpen
  \bibfield  {author} {\bibinfo {author} {\bibfnamefont {K.~W.}\ \bibnamefont
  {Bewig}}\ and\ \bibinfo {author} {\bibfnamefont {W.~A.}\ \bibnamefont
  {Zisman}},\ }\bibfield  {title} {\enquote {\bibinfo {title} {The wetting of
  gold and platinum by water},}\ }\href@noop {} {\bibfield  {journal} {\bibinfo
   {journal} {J. Phys. Chem.}\ }\textbf {\bibinfo {volume} {69}},\ \bibinfo
  {pages} {4238--4242} (\bibinfo {year} {1965})}\BibitemShut {NoStop}%
\bibitem [{\citenamefont {Huzayyin}\ \emph {et~al.}(2014)\citenamefont
  {Huzayyin}, \citenamefont {Chang}, \citenamefont {Lian},\ and\ \citenamefont
  {Dawson}}]{huzayyin14}%
  \BibitemOpen
  \bibfield  {author} {\bibinfo {author} {\bibfnamefont {A.}~\bibnamefont
  {Huzayyin}}, \bibinfo {author} {\bibfnamefont {J.~H.}\ \bibnamefont {Chang}},
  \bibinfo {author} {\bibfnamefont {K.}~\bibnamefont {Lian}}, \ and\ \bibinfo
  {author} {\bibfnamefont {F.}~\bibnamefont {Dawson}},\ }\bibfield  {title}
  {\enquote {\bibinfo {title} {{Interaction of water molecules with Au(111) and
  Au(110) surfaces under the influence of an external electric field}},}\
  }\href@noop {} {\bibfield  {journal} {\bibinfo  {journal} {J. Phys. Chem. C}\
  }\textbf {\bibinfo {volume} {118}},\ \bibinfo {pages} {3459--3470} (\bibinfo
  {year} {2014})}\BibitemShut {NoStop}%
\bibitem [{\citenamefont {Velasco-Velez}\ \emph {et~al.}(2014)\citenamefont
  {Velasco-Velez}, \citenamefont {Wu}, \citenamefont {Pascal}, \citenamefont
  {Wan}, \citenamefont {Guo}, \citenamefont {Prendergast},\ and\ \citenamefont
  {Salmeron}}]{velasco14}%
  \BibitemOpen
  \bibfield  {author} {\bibinfo {author} {\bibfnamefont {J.-J.}\ \bibnamefont
  {Velasco-Velez}}, \bibinfo {author} {\bibfnamefont {C.~H.}\ \bibnamefont
  {Wu}}, \bibinfo {author} {\bibfnamefont {T.~A.}\ \bibnamefont {Pascal}},
  \bibinfo {author} {\bibfnamefont {L.~F.}\ \bibnamefont {Wan}}, \bibinfo
  {author} {\bibfnamefont {J.}~\bibnamefont {Guo}}, \bibinfo {author}
  {\bibfnamefont {D.}~\bibnamefont {Prendergast}}, \ and\ \bibinfo {author}
  {\bibfnamefont {M.}~\bibnamefont {Salmeron}},\ }\bibfield  {title} {\enquote
  {\bibinfo {title} {The structure of interfacial water on gold electrodes
  studied by x-ray absorption spectroscopy},}\ }\href@noop {} {\bibfield
  {journal} {\bibinfo  {journal} {Science}\ ,\ \bibinfo {pages} {1259437}}
  (\bibinfo {year} {2014})}\BibitemShut {NoStop}%
\bibitem [{\citenamefont {Armano}\ \emph
  {et~al.}(2018{\natexlab{b}})\citenamefont {Armano} \emph
  {et~al.}}]{armano18a}%
  \BibitemOpen
  \bibfield  {author} {\bibinfo {author} {\bibfnamefont {M.}~\bibnamefont
  {Armano}} \emph {et~al.},\ }\bibfield  {title} {\enquote {\bibinfo {title}
  {Beyond the required {LISA} free-fall performance: {New LISA Pathfinder}
  results down to $20\mu hz$},}\ }\href@noop {} {\bibfield  {journal} {\bibinfo
   {journal} {Phys. Rev. Letts.}\ }\textbf {\bibinfo {volume} {120}},\ \bibinfo
  {pages} {061101} (\bibinfo {year} {2018}{\natexlab{b}})}\BibitemShut
  {NoStop}%
\end{thebibliography}%

\end{document}